\documentclass[twocolumn,prb,showpacs,preprintnumbers,groupedaddress,amsmath,amssymb]{revtex4}
\usepackage{graphicx}
\usepackage{dcolumn}
\usepackage{color}
\usepackage{bm}
\numberwithin{equation}{section}

\DeclareMathOperator{\sgn}{sgn}

\DeclareMathOperator{\diag}{diag}

\newcommand{\br}{\mathbf{r}}
\newcommand{\ud}{\,\mathrm{d}}

\newcommand{\tphi}{\tilde{\phi}}
\newcommand{\tPi}{\tilde{\Pi}}
\newcommand{\ttheta}{\tilde{\theta}}
\newcommand{\im}{\textrm{i}}
\newcommand{\xir}{\xi_{\nu \beta \delta}^{r}}

\newcommand{\e}{\text{e}}
\newcommand{\be}{\begin{equation}}
\newcommand{\ee}{\end{equation}}
\newcommand{\bea}{\begin{eqnarray}}
\newcommand{\eea}{\end{eqnarray}}

\renewcommand{\vec}[1]{\boldsymbol #1}

\def\l{\left}
\def\r{\right}
\def\12{\frac{1}{2}}

\begin{document}

\title{The Two-Band Luttinger Liquid with Spin-Orbit Coupling:\\ Applications to Monatomic Chains on Surfaces}

\author{N. Sedlmayr}
\email{sedlmayr@physik.uni-kl.de}
\affiliation{Department of Physics and Research Center OPTIMAS,
  Technical University Kaiserslautern,
  D-67663 Kaiserslautern, Germany}
\author{P. Korell}
\affiliation{Department of Physics and Research Center OPTIMAS,
  Technical University Kaiserslautern,
  D-67663 Kaiserslautern, Germany}
\author{J. Sirker}
\affiliation{Department of Physics and Research Center OPTIMAS,
  Technical University Kaiserslautern,
  D-67663 Kaiserslautern, Germany}

\date{\today}

\begin{abstract}
  Recently, monatomic chains on surfaces have been synthesized which
  show evidence of Luttinger liquid physics. The experimental data
  point to a dispersion along the chain with four Fermi points. Here
  we investigate a general low-energy effective Hamiltonian for such a
  two-band model where $SU(2)$ spin symmetry is broken but time
  reversal symmetry persists, as is expected due to the surface
  geometry. Spin-orbit coupling gives rise to a new energy scale
  $\varepsilon_{\textrm{SO}}$ much smaller than the Fermi energy
  $\varepsilon_F$ and to spin non-conserving scattering processes. We
  derive the generic phase diagram at zero temperature as well
  as an effective phase diagram at temperatures
  $\varepsilon_{\textrm{SO}}<T\ll \varepsilon_F$. For the part of the
  phase diagram where a Luttinger liquid is found to be stable, the
  density of states and the spectral function are calculated and
  discussed in relation to the experimental data.
\end{abstract}


\maketitle

\section{Introduction}
Interacting one-dimensional itinerant electron systems behave very
differently from those in higher dimensions. The correct low-energy
theory to describe such systems is not Fermi liquid but rather
Luttinger liquid theory. In contrast to a Fermi liquid, a Luttinger
liquid has collective excitations, shows only a power law suppression
of the occupation number $n_k$ near the Fermi momentum $k_F$, i.e., a
zero quasiparticle weight, and a separation of spin and charge degrees
of freedom.\cite{Haldane1981a,Giamarchi2004}

Experimentally, a number of quasi one-dimensional structures have been
investigated with the aim to confirm Luttinger liquid behavior. This
has been particularly successful for spin chains, i.e., for systems
where the charge channel is gapped. Prominent examples are various
cuprate and organic chains with superexchange coupling constants along
the chain direction $J$ being orders of magnitude larger than the
interchain couplings $J_\perp$, making them to a very good
approximation one-dimensional at temperatures $J_\perp \ll T \ll
J$.\cite{MotoyamaEisaki,Hase1993a,Dender1997} For such systems it
has been possible to show that the Luttinger liquid {\it
  quantitatively} describes a large number of experimental data
ranging from thermodynamic measurements to dynamical response
functions.\cite{EggertAffleck92,egg94,OshikawaAffleck,SirkerLaflorencie,SirkerLaflorencie2,SirkerPereira,SirkerPereira2,PereiraSirker,PereiraSirkerJSTAT}

For itinerant electron systems mounting evidence has also been compiled
in recent years, verifying the predictions of Luttinger liquid
theory including, in particular, spin charge
separation.\cite{AuslaenderSteinberg,YaoPostma,Bockrath,Jompol2009,Deshpande2010}
The quasi one-dimensional systems these results have been obtained
for are, on the one hand, carbon nanotubes, and, on the other hand,
two-dimensional electron gases confined to a narrow channel by gate
electrodes.

Other possible candidates for Luttinger liquids are monatomic chains
on surfaces. The best studied example are gold chains on top of a
Si(111)
surface.\cite{SegoviaPurdie,Losio2001,Ahn2003,S'anchez-Portal2004,Barke2006}
While the gold chains were found to exhibit a metal-insulator
transition at an energy scale of $\sim 100$ K so that low-energy
Luttinger liquid physics could not be studied,\cite{Ahn2003,Barke2006}
a number of important general observations were nevertheless made. In
particular, angle resolved photo emission spectra (ARPES) have shown
two closely spaced bands which were first interpreted as a
signature of spin-charge separation.\cite{SegoviaPurdie} Later though it has
been shown by ab-initio calculations\cite{S'anchez-Portal2004}
and a more detailed ARPES study\cite{Barke2006} that the splitting of
the band is caused by spin-orbit coupling.

Very recently, a different surface system has been found which seems
to remain metallic down to temperatures of the order of a few
Kelvin.\cite{Blumenstein2011} Here Au atoms self organize into
chains on a Ge(001) surface and scanning tunneling spectroscopy (STS) has
revealed a density of states (DOS) showing power law scaling with
energy which is indicative of a possible Luttinger liquid state. A subsequent
ARPES study showed that the 1D character of the Au chains is indeed
exceptionally high. An additional complication in this system arises,
however, because the single surface band shows {\it two} electron
pockets,\cite{Schafer2008,Meyer2011} a fact which has to be taken into
account in a proper theoretical description.

In all these surface systems, Rashba and Dresselhaus-type spin-orbit
couplings are generically expected to be present due to the reduced
symmetry.\cite{Winkler2003,HoepfnerSchaefer} In particular, spin-rotational
symmetry is expected to be broken and only time-reversal symmetry will
persist. Luttinger liquids with spin-orbit interactions have been
studied previously in the context of carbon
nanotubes\cite{Schulz2010} and magnetized spin chains and
quantum wires.\cite{Gangadharaiah2008,Schulz2009} There is also a rather
extensive literature on one-dimensional models where two bands cross
the Fermi
surface.\cite{Varma1985,Penc1990,Finkel'stein1993,Fabrizio1993,Khveshchenko1994a,Balents1996,Tsuchiizu2002a,Tsuchiizu2005,Chudzinski2008,Noack1996,Khveshchenko1994,Sedlmayr2011b,Sedlmayr2012a,Sedlmayr2013}
However, in these works the bands are either assumed to have $SU(2)$
symmetry or to be completely spin split by a magnetic field.

In this paper we want to consider a generic two-band model with
spin-orbit coupling, including all interaction terms which are allowed
by time reversal symmetry. While in the $SU(2)$ symmetric case the
phase diagram is to a large extent determined by the renormalization
group (RG) flow of marginal interaction terms, most of these
interaction terms will become either relevant or irrelevant in the
case with only time reversal symmetry, simplifying the calculation of
the phase diagram. On the other hand, four instead of only two
independent Luttinger parameters are present once $SU(2)$ symmetry is
broken, leading to a much richer phase diagram. Of particular relevance
for the experiments on monatomic chains on surfaces is the question if
a Luttinger liquid phase can survive at all in a surface geometry
where the symmetries are reduced.  We will show that this is indeed the case,
however the phase diagram turns out to be quite rich.

This paper is organized as follows. In section \ref{sec_model} we
introduce a general two-band model with spin-orbit coupling, show how
to bosonize it, and look at the interaction terms which are allowed by
time reversal symmetry. In section \ref{sec_symmetry} we calculate the
spin density correlation functions and analyze the simplifications for
the specific point in parameter space where the model has an
additional $SU(2)$ symmetry. Section \ref{sec_phase} is devoted to a
renormalization group analysis and the subsequent phase diagram of the
general model. The spin-flip scattering terms present are found to be
slowly oscillating in space so that they can only be ignored at the
lowest temperatures. We therefore also present an effective phase
diagram at small temperatures where these terms are still present in
the RG flow. In section \ref{sec_spectral} we look at the spectral
function and density of states and try to determine in which part of
the phase diagram the system of monatomic chains might be located.
Finally, in section \ref{conclusions} we conclude.

\section{Model}\label{sec_model}

\subsection{The non-interacting band structure}\label{llmodel}

We are interested in wires composed of single atoms deposited in
surfaces. Such wires have no structural inversion symmetry or
inversion centre and therefore both Rashba and Dresselhaus-type
couplings can be present.\cite{HoepfnerSchaefer} For small spin-orbit
coupling this will not lead to any drastic effects for the
noninteracting band structure.  Nonetheless the breaking of $SU(2)$
symmetry does have important consequences for the interaction
terms.\cite{Giamarchi1988}

In experiment, the surface band is found to cross the Fermi energy
four times, forming two small electron pockets. No microscopic model
for the non-interacting band structure, i.e., by downfolding starting
from a density functional theory calculation, has been obtained yet.
The origin of this band structure is, however, not important for the
low-energy effective theory we are going to construct and we simply take as a given
the non-interacting Hamiltonian
\begin{equation}\label{genhamiltonian}
 H_0 = \sum_{\sigma k} \left( \epsilon_k - \mu \right) c_{\sigma k}^\dagger c_{\sigma k}
\end{equation}
where $\epsilon_k$ is the dispersion, $\mu$ the chemical potential,
and $c_{\sigma k}^{(\dagger)}$ the fermionic annihilation (creation)
operators for particles with spin $\sigma$ and momentum $k$. The dispersion before taking spin-orbit interactions into
account is supposed to have four spin degenerate Fermi points $\pm
k_{F 1}, \pm k_{F 2}$ in the Brillouin zone, i.e., $\epsilon_{k_{F b}}
= \mu$ for the ``bands'' $b = 1,2$. We label the Fermi momenta such
that $k_{F1}<k_{F2}$.

We are interested in the regime where the Fermi energy is much larger than the temperature and the
energy scale the response of the system is tested at experimentally. Therefore we can linearize the dispersion
at these Fermi points and introduce two Fermi velocities
\begin{equation}
 v_{F b} = \left\lvert\frac{\textup{d} \epsilon}{\textup{d} k} \right\rvert_{k = k_{F b}}\,.
\end{equation}
For clarity we shall later mostly focus on the situation where the
bands are completely symmetric such that $v_{F1}=v_{F2}$ and where the
density-density interactions are also band independent. The more
general case can be treated in a similar way and the necessary
generalization is shown in Appendix \ref{appendix_rotations}.

Through the standard linearization procedure in the vicinity of each
Fermi point new fermionic annihilation and creation operators,
$c_{\sigma r b} (q)$ and $c^\dagger_{\sigma r b} (q)$, can be defined
for particles in band $b$ with spin $\sigma=\pm$ and relative momentum $q
= k - r \eta_b k_{F b}$. Here $r = \pm$ indicates the direction in
which the particle is moving, and $\eta_b = (-1)^b$ is an additional band
factor depending on whether the slope of the dispersion $\epsilon_k$
at the Fermi point $+k_{F b}$ is positive or negative.

More precisely, we can make the following ansatz using a continuum
representation in position space
\begin{equation}\label{ansatz}
\psi_{\sigma} (x) = \sum_{br}\e^{\im r\eta_bk_{Fb}x}\psi_{\sigma r b}(x)\,,
\end{equation}
where the fields are given by the Fourier transformation
\begin{equation}
\sqrt{a}\psi_{\sigma r b}(x)=\frac{1}{L}\sum_q\e^{\im qx}c_{\sigma r b} (q)\,,
\end{equation}
with $a$ the lattice spacing.

\begin{figure}
\includegraphics[width=0.45\textwidth]{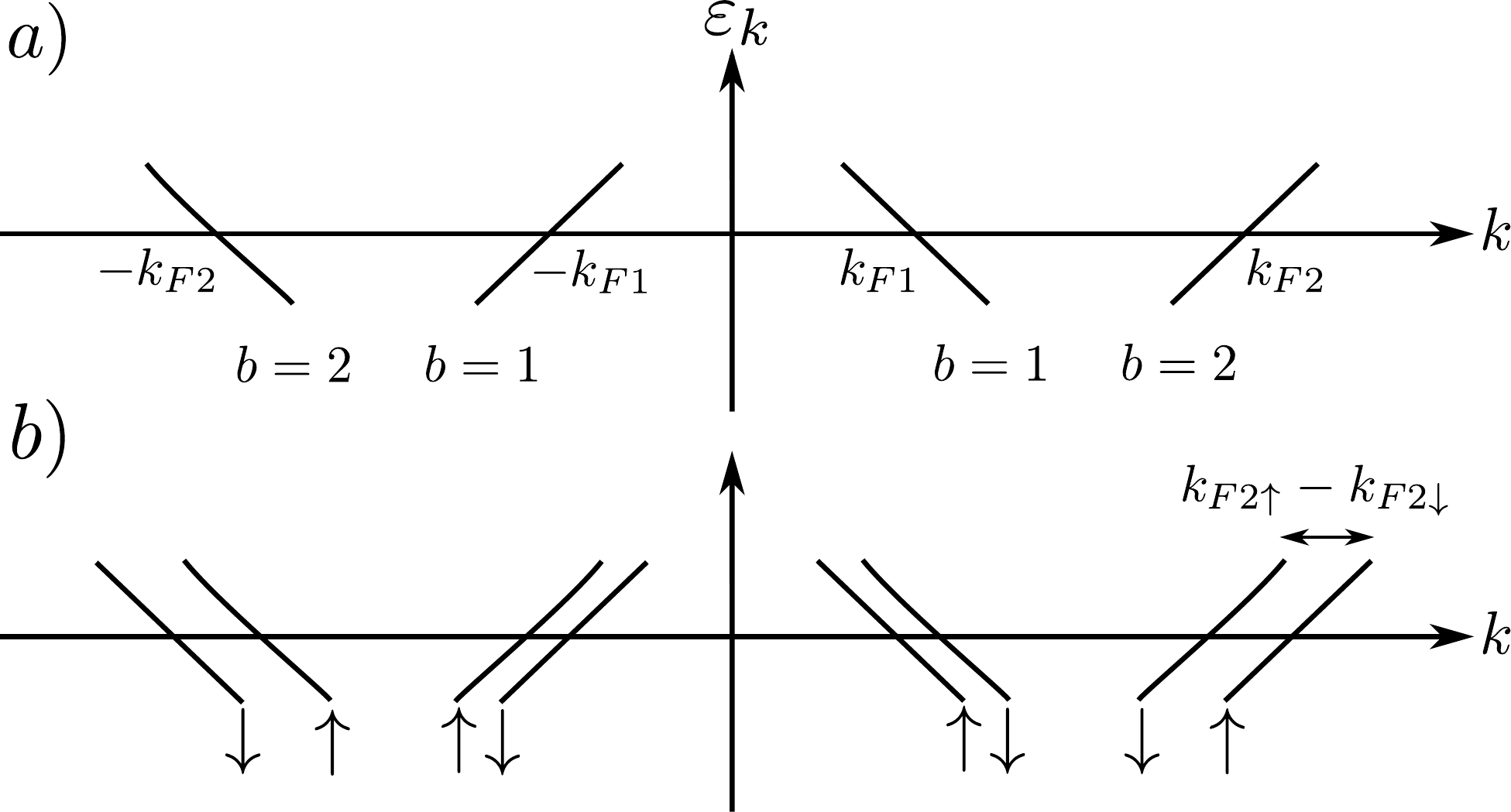}
\caption[Band Structure]{The dispersion around the Fermi points with
  the effective band indices $1,2$ which are used in the Hamiltonian
  (\ref{H0}).  Panel a) shows the bands without spin-orbit splitting,
  panel b) the spin-orbit split bands.}
\label{bands}
\end{figure}

Following this the Hamiltonian \eqref{genhamiltonian} becomes
\begin{equation}
\label{H0}
 H_0 = \sum_{\sigma r b} r v_{F b} \int \ud x \psi_{\sigma r b}^\dagger (x) \left( -\im \partial_x \right) \psi_{\sigma r b} (x)\,,
\end{equation}
a one-dimensional Dirac Hamiltonian with branches labeled by $(\sigma
r b)$, see Fig.~\ref{bands} a).

\subsection{Spin-orbit interactions}
\label{Sec_SO}
In order to properly treat the spin-orbit interaction one has to start from a two-dimensional Hamiltonian
\begin{eqnarray}
 H_{2D} &=& \int \ud x \ud y\; \vec{\psi}^\dagger(x,y) [\hat\epsilon_x + \hat\epsilon_y +V_c(y) \\
&+& \alpha (\hat p_x\sigma_y -\hat{p}_y \sigma_x) + \beta (\hat p_x\sigma_x -\hat{p}_y \sigma_y)]\vec{\psi}(x,y) \nonumber
\end{eqnarray}
where $\vec{\psi}=(\psi_\uparrow,\psi_\downarrow)$ and
$\hat{\epsilon}_i$ is the kinetic energy operator in the $i=x,y$
direction, $x$ being longitudinal along the wire. $V_c(y)$ is a
confining potential in the transverse direction. The terms in $\alpha$
and $\beta$ are the Rashba and Dresselhaus like spin-orbit coupling terms
respectively. The part $\sim -\hat p_y(\alpha\sigma_x + \beta
\sigma_y)$ can be treated perturbatively and leads to an effective
mixing of higher lying states into the lowest band. As a consequence,
the velocities of the spin-orbit split bands, see Fig.~\ref{bands}(b),
can become unequal.\cite{Schulz2010} For a strong confinement $V_c(y)$
this is a small effect which we will neglect in the following. Using
this approximation, we are left with a one-dimensional Hamiltonian
which can be diagonalized with a spin-orbit induced splitting of the
bands given by
\begin{equation}
\label{HSO}
H_{SO} =\int \ud x\sum_{r\sigma b} \eta_b r\sigma k_{Fb}\sqrt{\alpha^2+\beta^2}\psi^\dagger_{\sigma br}\psi_{\sigma br}.
\end{equation}
This band splitting gives rise to the definition of a new energy scale
$\varepsilon^b_{\textrm{SO}}\equiv v_{Fb}k^b_{\textrm{SO}}=\sqrt{\alpha^2+\beta^2} k_{Fb}$ which is much
smaller than the Fermi energy $\varepsilon_F$. Here we have assumed that the spin-orbit
couplings $\alpha$ and $\beta$ are the same for both bands.

\subsection{Quartic interaction terms allowed by time reversal
  symmetry}
A short range density-density interaction introduces terms which are
quadratic in terms of the densities $\rho_{\sigma
  br}=\psi^\dagger_{\sigma br}\psi_{\sigma br}$ as well as quartic
terms in the fermionic operators $\psi_{\sigma br}$ which cannot be
written as functions of the densities $\rho_{\sigma br}$. The possible
interaction processes depend on the symmetries of the underlying
microscopic model. Although the Coulomb interaction itself is $SU(2)$
symmetric, due to the spin-orbit interactions in Eq.~(\ref{HSO})
scattering between electrons with the same or different spins are no
longer equivalent.\cite{Giamarchi1988} Furthermore, spin-flip
scattering processes become possible. We consider here the most
general form of interactions for a system where only time reversal
symmetry is present. Later, we make two simplifications: (1) We take
the velocities of the spin split bands as being equal,
$v_{Fb\sigma}=v_{Fb\bar\sigma}$. As explained above, this is
expected to be an excellent approximation for systems where the
confining potential $V_c(y)$ is strong. (2) We will mainly focus on
the case when the bands are also symmetric, $v_{Fb\sigma}=v_{F\bar b
  \sigma}$, the generalization to nonsymmetric bands (though still
with $k\to-k$ symmetry), however, is straightforward and given in
Appendix \ref{appendix_rotations}.

First we introduce the notation we use to label the different inter-
and intra-band scattering processes. We retain the usual $g_1,g_2,g_4$
notation (`g-ology') for the electron directions. Therefore, as usual
$g_1$ refers to processes where the incoming electrons have different
directions and both backscatter.  $g_2$ refers to processes where the
incoming electrons have different directions and there is no
backscattering.
$g_4$ refers to incoming electrons of the same direction which do not
backscatter. In addition we use $\bar{g}$ to mean that the incoming
electrons are on different bands and $g'$ to mean that the band index
is changed for both electrons.  Processes where one band index changes
and the other does not do not contribute (except for one umklapp
process treated separately in Appendix \ref{appendix_umklapp}).
Finally we have the spin degrees of freedom. $g_\parallel$ and $g_\perp$ always refer to the spin indices
and denote a process where the electrons have the same spin or
different spins respectively.  Where there is no ambiguity and both $g_\parallel$ and $g_\perp$
terms are present these indices will sometimes be suppressed. $g_s$ is used to refer to a process where
two up spins are scattered to two down spins or vice versa.

\subsubsection{Density-density type interactions}\label{d-d}
Exactly as in the single band Luttinger liquid there are $g_2$ and
$g_4$ density-density interactions, but these no longer need to lie on
the same band, thus
\begin{eqnarray}
H_2&=&\sum_{\sigma\sigma'br}\int\ud x\left[\frac{g_{2b}}{2}\rho_{\sigma' b \bar{r}}\rho_{\sigma b r}+ \frac{\bar{g}_2}{2}\rho_{\sigma'\bar{b} \bar{r}}\rho_{\sigma b r}\right]\nonumber\,,\\
H_4&=&\sum_{\sigma\sigma'br}\int\ud x\left[\frac{g_{4b}}{2}\rho_{\sigma' b r}\rho_{\sigma b r}+ \frac{\bar{g}_4}{2}\rho_{\sigma' \bar{b} r}\rho_{\sigma b r}\right]\,.\qquad
\end{eqnarray}
There are also several backscattering terms which can be rearranged
into density-density interactions in the standard way,
\begin{eqnarray}
H_{1\parallel}&=&-\sum_{\sigma br}\int\ud x\frac{g_{1\parallel}}{2}\rho_{\sigma b \bar{r}}\rho_{\sigma b r}\,,\nonumber\\
\bar{H}'_{1\parallel}&=&-\sum_{\sigma br}\int\ud x\frac{\bar{g}'_{1\parallel}}{2}\rho_{\sigma \bar{b} \bar{r}}\rho_{\sigma b r}\,,\\
\bar{H}'_{4\parallel}&=&-\sum_{\sigma br}\int\ud x\frac{\bar{g}'_{4\parallel}}{2}\rho_{\sigma \bar{b} r}\rho_{\sigma b r}\,,\nonumber
\end{eqnarray}
and rescale the $g_{2\parallel}$, $\bar{g}_{2\parallel}$ and
$\bar{g}_{4\parallel}$ interactions, respectively. These three terms will be assumed to have already
been incorporated into their kinematic equivalents, and will not be made explicit in the following.
The same will be done for all other equivalent processes we find.

\subsubsection{Backscattering and inter-band scattering}
With the addition of an extra band many more backscattering, inter-band scattering, and
umklapp processes become possible, which have no equivalent for a
single band model. As a consequence, we might expect that the extent
of a Luttinger liquid phase in the phase diagram---if it survives at
all---will be much smaller than in a single band model.

\begin{figure}
\includegraphics[width=0.45\textwidth]{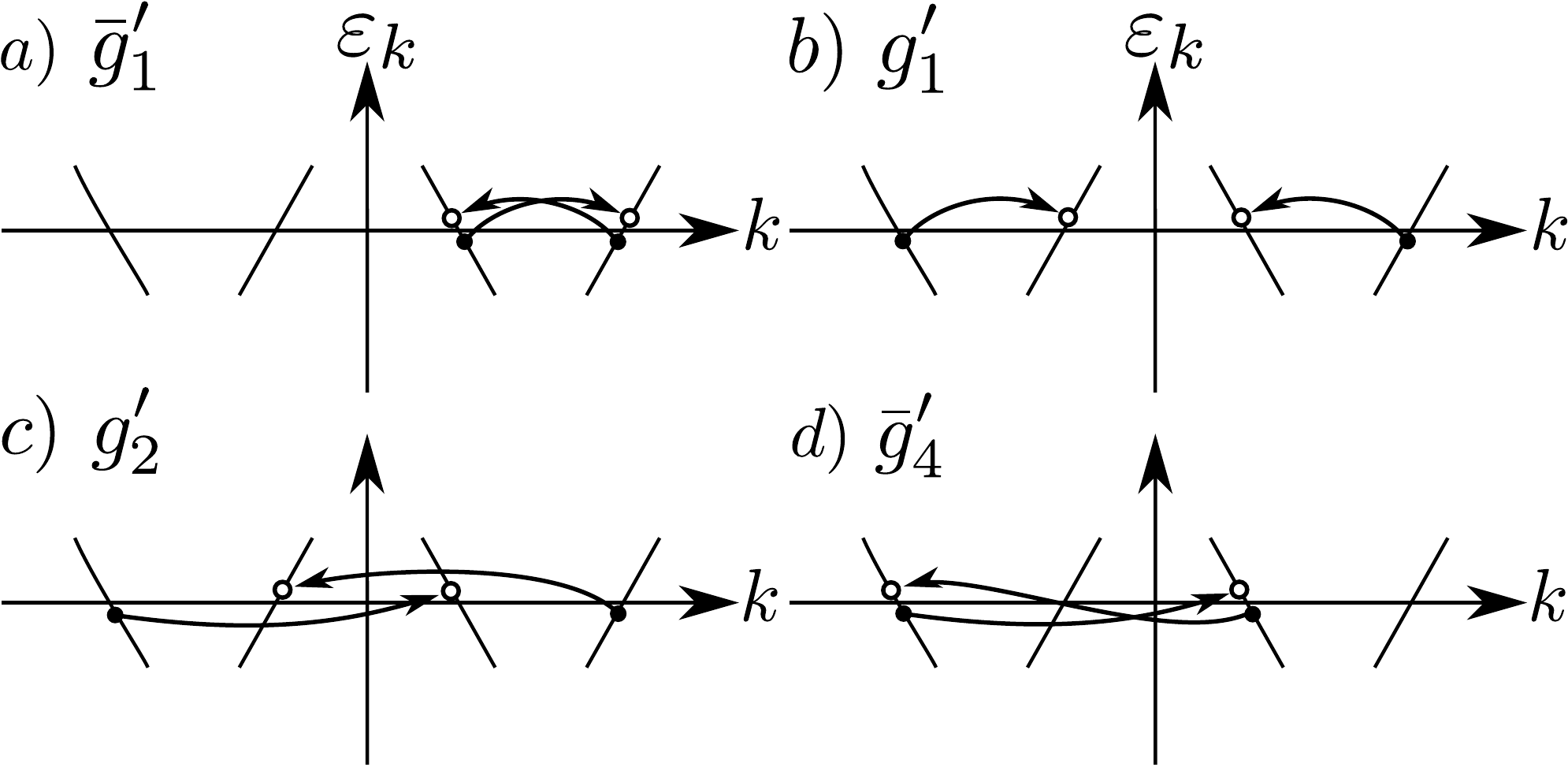}
\caption[Scattering Terms]{Additional zero momentum transfer backscattering and inter-band
scattering processes in a two-band model, see Eq.~\eqref{int_bs}. Band structure as for Fig.~\ref{bands} a).}
\label{scattering}
\end{figure}

We confine ourselves here to the zero momentum transfer terms, with
all other terms suppressed by rapid oscillations in the integrals.
Some additional umklapp scattering and backscattering processes which become
non-oscillating at special commensurate fillings and thus do
contribute at these special fillings are treated in Appendix
\ref{appendix_umklapp}. The generic non-oscillating backscattering
terms are
\begin{eqnarray}\label{int_bs}
H_{1\perp}&=&\sum_{\sigma br}\int\ud x\frac{g_{1\perp}}{2}\psi^\dagger_{\bar{\sigma} b r}\psi_{\bar{\sigma} b \bar{r}}\psi^\dagger_{\sigma b\bar{r}}\psi_{\sigma br}\,,\nonumber\\
\bar{H}'_{1\perp}&=&\sum_{\sigma br}\int\ud x\frac{\bar{g}'_{1\perp}}{2}\psi^\dagger_{\bar{\sigma} br}\psi_{\bar{\sigma} \bar{b} \bar{r}} \psi^\dagger_{\sigma\bar{b}\bar{r}}\psi_{\sigma br}\,,\nonumber\\
H'_{1}&=&\sum_{\sigma,\sigma',b,r}\int\ud x\frac{g'_{1}}{2}\psi^\dagger_{\sigma' \bar{b} r}\psi_{\sigma' b \bar{r}}\psi^\dagger_{\sigma\bar{b}\bar{r}}\psi_{\sigma br}\,,\\
H'_{2\perp}&=&\sum_{\sigma,b,r}\int\ud x\frac{g'_{2\perp}}{2}\psi^\dagger_{\bar{\sigma} \bar{b} \bar{r}}\psi_{\bar{\sigma} b \bar{r}}\psi^\dagger_{\sigma\bar{b}r}\psi_{\sigma br}\nonumber\,\textrm{, and}\\
\bar{H}'_{4\perp}&=&\sum_{\sigma br}\int\ud x\frac{\bar{g}'_{4\perp}}{2}\psi^\dagger_{\bar{\sigma} b r}\psi_{\bar{\sigma}\bar{b}r}\psi^\dagger_{\sigma\bar{b}r}\psi_{\sigma br}\,.\nonumber
\end{eqnarray}
The additional backscattering and inter-band processes possible in a
two-band, as opposed to a single band, model are shown
schematically in Fig.~\ref{scattering}.

\subsubsection{Single spin-flip scattering}
So far, we have just considered the generalization of the usual
Coulomb interaction terms to the case of a two-band model. All these
terms are symmetric under time reversal
\begin{equation}
\label{time_reverse}
\Psi_{\sigma br}\to \sigma \Psi^\dagger_{\bar\sigma b\bar r}
\end{equation}
However, there are also additional spin-flip scattering processes
allowed by time reversal symmetry.

The first cases we want to look at are those in which a single spin is
flipped during the scattering process. For every possible
$g_\parallel$ and $g_\perp$ scattering process listed in the two
preceding sections, such a single spin flip term can be constructed.
For example, a $g_1$-type single spin-flip process has the form $\sim
\psi^\dagger_{\sigma' b r}\psi_{\sigma'
  b\bar{r}}\psi^\dagger_{\bar{\sigma}b\bar{r}}\psi_{\sigma br}.$ It
is, however, easy to check that this term is always irrelevant. There
are a number of similar terms which turn out to be irrelevant as well.
The only two single spin-flip terms which, in principle, can become
relevant, see Eq.~\eqref{spin-flip_1}, are
\begin{eqnarray}\label{relsf}
H'_{1sf}&=&\frac{G_1}{2}\sum_{\sigma,\sigma',b,r}r\sigma\sigma'\!\int\!\ud x\, \psi^\dagger_{\sigma' \bar{b} r}\psi_{\sigma' b \bar{r}}\psi^\dagger_{\bar{\sigma}\bar{b}\bar{r}}\psi_{\sigma br}\nonumber\textrm{, and }\,\\
H'_{2sf}&=&\frac{G_2}{2}\sum_{\sigma,\sigma',b,r}r\sigma\sigma'\!\int\!\ud x\,  \psi^\dagger_{\sigma' \bar{b} \bar{r}}\psi_{\sigma' b \bar{r}}\psi^\dagger_{\bar{\sigma}\bar{b}r}\psi_{\sigma br}.
\end{eqnarray}

It is important to note that all the single spin-flip terms are slowly
oscillating due to the band splitting caused by the spin-orbit
coupling, see Fig.~\ref{bands}(b). They can therefore only affect the
behavior of the model at intermediate temperatures while we have to
drop them anyways at zero temperature. Furthermore, single spin-flip
terms are completely forbidden if the velocities of the spin-split
bands are equal, $v_{Fb\sigma}=\ v_{Fb\bar\sigma}$. In this case,
$\sigma\to\bar\sigma$ and $r\to \bar{r}$ are, up to the appropriate
shifts in momentum, separate symmetries of the Hamiltonian. As
discussed in Sec.~\ref{Sec_SO}, the velocities do become unequal once
the mixing with other transverse modes is taken into account. We have
assumed this to be a small effect and neglect the two scattering terms
\eqref{relsf} completely in the following.

\subsubsection{Double spin-flip scattering}
\begin{figure}
\includegraphics[width=0.45\textwidth]{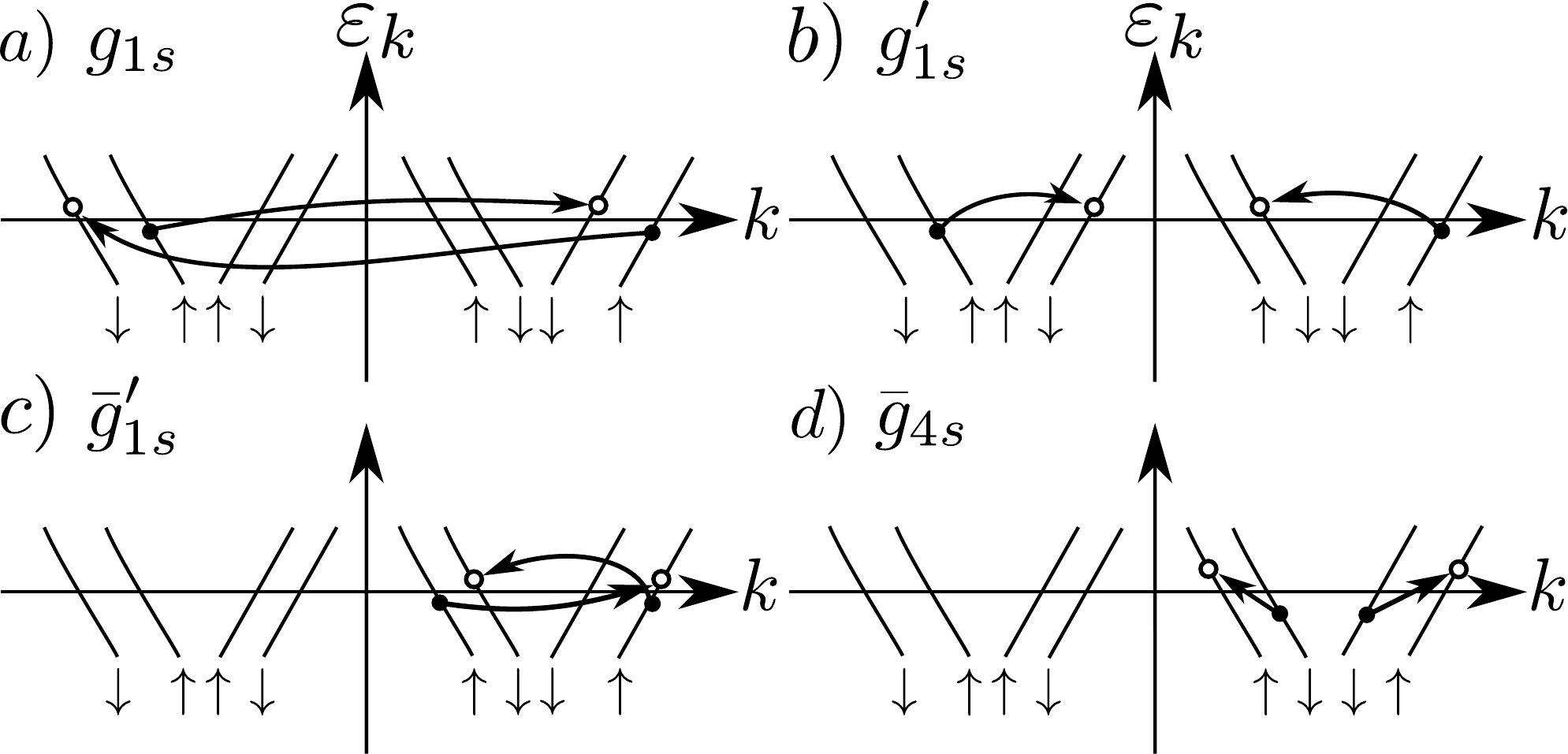}
\caption[Scattering Terms]{Double spin-flip processes allowed in a
  two-band model with time reversal symmetry, see Eq.~\eqref{int_dsf0}
  and Eq.~\eqref{int_dsf}. Band structure as for Fig.~\ref{bands} b), with the spin
splitting due to spin orbit coupling made explicit.}
\label{Fig_spin-flip}
\end{figure}

The other class of additional scattering terms allowed by time
reversal symmetry, Eq.~(\ref{time_reverse}), are double spin-flip
processes. As for single-flip scattering we can, starting from the
usual backscattering, inter-band scattering, and density-density terms, construct all
possible double spin-flip
processes. These contributions are
\begin{eqnarray}
\label{int_dsf0}
H_{1s}=\frac{g_{1s}}{2}\sum_{\sigma br}\int\ud x\;\psi^\dagger_{\bar{\sigma} b r}\psi_{\sigma b \bar{r}}\psi^\dagger_{\bar{\sigma} b\bar{r}}\psi_{\sigma br}\,,\nonumber\\
H'_{1s}=\frac{g'_{1s}}{2}\sum_{\sigma br}\int\ud x\;\psi^\dagger_{\bar{\sigma} \bar{b} r}\psi_{\sigma b \bar{r}}\psi^\dagger_{\bar{\sigma} \bar{b}\bar{r}}\psi_{\sigma br}\,,\\
\bar H'_{1s}=\frac{\bar{g}'_{1s}}{2}\sum_{\sigma br}\int\ud x\;\psi^\dagger_{\bar{\sigma} b r}\psi_{\sigma \bar{b} \bar{r}}\psi^\dagger_{\bar{\sigma} \bar{b}\bar{r}}\psi_{\sigma br}\,.\nonumber
\end{eqnarray}
Three $g_2$ processes also exist.  However they are not kinematically
distinct from the $g_{1s}$ interactions. For completeness they would
be $g_{2s}$, $g'_{2s}$ and $\bar{g}_{2s}$ and are equivalent to
$g_{1s}$, $g'_{1s}$ and $\bar{g}'_{1s}$ respectively. In addition,
there are two kinematically distinct $g_4$ processes ($\bar{g}'_{4s}$ is equivalent to $\bar{g}_{4s}$):
\begin{eqnarray}\label{int_dsf}
H_{4s}=\frac{g_{4s}}{2}\sum_{\sigma br}\int\ud x\;\psi^\dagger_{\bar{\sigma} b r}\psi_{\sigma b r}\psi^\dagger_{\bar{\sigma} br}\psi_{\sigma br}\,,\nonumber\\
\bar H_{4s}=\frac{\bar{g}_{4s}}{2}\sum_{\sigma br}\int\ud x\;\psi^\dagger_{\bar{\sigma}\bar{b}r}\psi_{\sigma \bar{b}r}\psi^\dagger_{\bar{\sigma} br}\psi_{\sigma br}\,.
\end{eqnarray}
The double spin-flip scattering processes are shown schematically in
Fig.~\ref{Fig_spin-flip}.

Overall, the possible interaction processes consist of the
density-density type interactions, the backscattering and inter-band
terms, and the double spin-flip terms. Using bosonization, the
density-density terms can be absorbed into the quadratic Luttinger
liquid Hamiltonian while the backscattering, inter-band, and double
spin-flip terms will lead to interactions between the bosons. We will
see that the spin-orbit induced splitting of the bands,
Eq.~(\ref{HSO}), can also be absorbed into the Luttinger liquid
Hamiltonian by a shift in the bosonic fields. As a consequence,
however, some of the backscattering and all double spin-flip terms will
become slowly oscillating in space.  The full details are worked
out in the next section.

\subsection{Bosonization}
To bosonize the two-band model we introduce bosonic fields
$\phi_{\sigma r b} (x)$ for each branch\cite{Haldane1981a,Giamarchi2004}
\begin{equation}\label{vertex}
 \psi_{\sigma r b} (x) = \frac{1}{\sqrt{2 \pi \alpha}} \textup{e}^{\im  r \sqrt{2\pi}\phi_{\sigma r b} (x)}
\end{equation}
where the bosonic fields satisfy the
following commutation relations
\begin{equation}
\left[ \phi_{\sigma r b} (x), \phi_{\sigma' r' b'} (x') \right] = \delta_{\sigma \sigma'} \delta_{r r'} \delta_{b b'} \frac{\im r}{2} \sgn \left( x - x' \right)\,.
\end{equation}
$\alpha$ is a short distance cutoff of the order of the lattice
spacing $a$.  For the density operators $\rho_{\sigma r b} =
\psi_{\sigma r b}^\dagger \psi_{\sigma r b}$ this leads to the
expression $\sqrt{2 \pi} \rho_{\sigma r b} (x) = -\partial_x
\phi_{\sigma r b} (x)$.  The Hamiltonian (\ref{H0}) of the
non-interacting system can now be written in terms of the bosonic
fields
\begin{equation}
\label{H0_2}
 H_0 = \sum_{\sigma r b} \frac{v_{F b}}{2} \int \ud x \left( \partial_x \phi_{\sigma r b} (x) \right)^2\,.
\end{equation}
Similarly, the spin-orbit term (\ref{HSO}) can be bosonized, leading to
\begin{equation}
\label{HSO_2}
  H_{\textrm{SO}}=-\frac{\sqrt{\alpha^2+\beta^2}}{2\pi}\sum_{\sigma br} \eta_b r\sigma k_{Fb} \int \ud x \;\partial_x\phi_{\sigma br}\,.
\end{equation}

We can now also add the density-density type interactions from section
\ref{d-d} which are quadratic in the bosonic fields. The Hamiltonian
(\ref{H0_2}) including these interaction terms can be written as
\begin{eqnarray}\label{quadratic_hamiltonian}
H_q&=&\int\ud z[\partial_x\Phi(x)]^T\mathbf{M}\partial_x\Phi(x)\,,
\end{eqnarray}
where
\begin{eqnarray}\nonumber
 [\Phi]^T = (\phi_{\uparrow 1+},\phi_{\uparrow 1-},\phi_{\downarrow 1+},\phi_{\downarrow 1-}, \phi_{\uparrow 2+},\phi_{\uparrow 2-},\phi_{\downarrow 2+},\phi_{\downarrow 2-})
\end{eqnarray}
and $\mathbf{M}$ is a symmetric $8 \times 8$ matrix.  The bosonization
procedure is thus sufficient to re-express all but a few contributions
in terms of a diagonalizable quadratic bosonic Hamiltonian. The
matrix, $\mathbf{M}$, can be written as
\begin{equation}\label{m1m2}
\mathbf{M}=\frac{1}{2}\begin{pmatrix}
\mathbf{M}_1 & \mathbf{M}'' \\
\mathbf{M}'' & \mathbf{M}_2
\end{pmatrix}\,,
\end{equation}
where
\begin{equation}
\mathbf{M}_b=\begin{pmatrix}
v_{Fb}+\frac{g_{4\parallel b}}{2\pi} & \frac{g_{2\parallel b}}{4\pi} & \frac{g_{4\perp b}}{4\pi} & \frac{g_{2\perp b}}{8\pi} \\
\frac{g_{2\parallel b}}{4\pi} & v_{Fb}+\frac{g_{4\parallel b}}{2\pi} & \frac{g_{2\perp b}}{8\pi} & \frac{g_{4\perp b}}{4\pi} \\
\frac{g_{4\perp b}}{4\pi} & \frac{g_{2\perp b}}{8\pi} & v_{Fb}+\frac{g_{4\parallel b}}{2\pi} & \frac{g_{2\parallel b}}{4\pi}  \\
\frac{g_{2\perp b}}{8\pi} & \frac{g_{4\perp b}}{4\pi} & \frac{g_{2\parallel b}}{4\pi} & v_{Fb}+\frac{g_{4\parallel b}}{2\pi}
\end{pmatrix}\,,
\end{equation}
with $b=1,2$ labeling the bands, and
\begin{equation}
\mathbf{M}''=\frac{1}{8\pi}\begin{pmatrix}
2\bar{g}_{4\parallel }& 2\bar{g}_{2\parallel } & \bar{g}_{4\perp} & \bar{g}_{2\perp} \\
2\bar{g}_{2\parallel } & 2\bar{g}_{4\parallel}& \bar{g}_{2\perp} & \bar{g}_{4\perp} \\
\bar{g}_{4\perp} & \bar{g}_{2\perp} & 2\bar{g}_{4\parallel}& 2\bar{g}_{2\parallel } \\
\bar{g}_{2\perp} & \bar{g}_{4\perp} & 2\bar{g}_{2\parallel } & 2\bar{g}_{4\parallel}
\end{pmatrix}\,.
\end{equation}
$\mathbf{M}$ is a real symmetric matrix with real eigenvalues and
we can now diagonalize $\mathbf{M}$. The diagonalization can be split
up into several steps, and we will show the full procedure to make
connections with the standard single band Luttinger liquid clear.

To begin we make two unitary transformations. The first is $\phi_{\sigma
  b\pm}(x)=[\phi_{\sigma b}(x)\mp \theta_{\sigma b}(x)]/\sqrt{2}$.
Note that $\theta_{\sigma b}$ is the adjoint of $\phi_{\sigma b}$ and
they satisfy $[\phi_{\sigma b}(x),\Pi_{\sigma b}(x')]=i\delta(x-x')$
where $\Pi_{\sigma b}(x)=\partial_x\theta_{\sigma b}(x)$. This first
rotation has the effect of uncoupling the two adjoint fields. The
second transformation is to rotate to the spin-charge representation:
$\phi_{c/s,b}(x)=[\phi_{\uparrow b}(x)\pm\phi_{\downarrow
  b}(x)]/\sqrt{2}$ (and similarly for the $\theta(x)$ fields). The
effect of these two rotations can be summarized as
$\mathbf{M}'=\tilde{\mathcal{U}}^{-1}\mathcal{U}^{-1}\mathbf{M}\mathcal{U}\tilde{\mathcal{U}}$
with $[\Phi'(x)]^T=[\Phi(x)]^T\mathcal{U}\tilde{\mathcal{U}}$. Thus far this just corresponds to the
diagonalization procedure for the usual Luttinger liquid
\cite{Giamarchi2004} applied to the two bands separately.

We now have
\begin{equation}
[\Phi'(x)]^T=(\phi_{s1},\phi_{s2},\theta_{s1},\theta_{s2},\phi_{c1},\phi_{c2}
\theta_{c1},\theta_{c2})
\end{equation}
and the rotated Hamiltonian is defined by the diagonal matrix
\begin{equation}
\mathbf{M}'=\diag\l[\mathbf{M}'_{\phi s}, \mathbf{M}'_{\theta s}, \mathbf{M}'_{\phi c}, \mathbf{M}'_{\theta c}\r]\,.
\end{equation}
In the following we focus on the symmetric band case where
$\mathbf{M}_1=\mathbf{M}_2$, see Eq.~\eqref{m1m2}.  All results are straight forwardly
generalizable to the asymmetric case, and the appropriate formulae are
given in Appendix \ref{appendix_rotations}. For the spin and charge ($\nu=s,c$) sectors, the matrix blocks of $\mathbf{M}'$ are then
\begin{equation}
\mathbf{M}'_{\phi \nu}=\frac{1}{2}\begin{pmatrix}
\frac{v_{\nu}}{K_{\nu}} & v_{\nu B}  \\
v_{\nu B} & \frac{v_{\nu}}{K_{\nu}} \\
\end{pmatrix},
\end{equation}
and
\begin{equation}
\mathbf{M}'_{\theta \nu}=\frac{1}{2}\begin{pmatrix}
 v_{\nu} K_{\nu} & v_{\nu A} \\ v_{\nu A} & v_{\nu} K_{\nu}
\end{pmatrix}.
\end{equation}
Here $K_{s}$ and $K_{c}$ are the spin and charge Luttinger parameters
for the two bands, and $v_{s}$ and $v_{c}$ are the spin and charge
velocities. The off-diagonal parameters
$\{v_{sA},v_{sB},v_{cA},v_{cB}\}$ describe the coupling between the
fields for different bands in the spin and charge sectors and are
functions of the various $g_2$ and $g_4$ interaction parameters. Their
explicit form is given in Eqs.~(\ref{A9}, \ref{A10}) in Appendix \ref{app_param}.

In the band symmetric case we can simply
perform another rotation to diagonalize $\mathbf{M}'$ given by
$\phi_{\nu 1,2}(x)=[\phi^{\nu +}(x)\mp \phi^{\nu -}(x)]/\sqrt{2}$ and
$\theta_{\nu 1,2}(x)=[\theta^{\nu +}(x)\mp \theta^{\nu-}(x)]/\sqrt{2}$.
After these additional rotations the quadratic Hamiltonian becomes diagonal for symmetric bands,
\begin{equation}
\label{H_quad}
H_q = \sum_{\substack{\nu = c,s\\\beta=\pm}}\frac{u^{\nu \beta}K^{\nu\beta}}{2}
\int \ud x \left[ \frac{( \partial_x \phi^{\nu \beta} (x))^2}{\left(K^{\nu\beta}\right)^2} + ( \Pi^{\nu \beta} (x))^2 \right]\,,
\end{equation}
where $u^{\nu \beta}$ ($K^{\nu\beta}$) are renormalized velocities
(Luttinger parameters), the conjugate momenta are given by $\Pi^{\nu
\beta} (x) = \partial_x \theta^{\nu \beta} (x)$ and the fields obey
bosonic commutation relations
\begin{equation}
 \left[ \phi^{\nu \beta} (x), \Pi^{\nu' \beta'} (x') \right] = \im \delta_{\nu \nu'} \delta_{\beta \beta'} \delta \left( x - x' \right)\,.
\end{equation}
$\beta=\pm$ label symmetric and antisymmetric combinations of the
bands, analogous to charge and spin in the spin subspace.

Finally, we can also rewrite the spin-orbit Hamiltonian (\ref{HSO_2})
in terms of the new fields
\begin{equation}
H_{\textrm{SO}}=-\frac{\sqrt{2(\alpha^2+\beta^2)}}{\pi}\int d x\, [k_F\partial_x\theta^{s-}+\tilde{k}\partial_x\theta^{s+}]\,,
\end{equation}
with $k_{Fb} = k_F + \eta_b \tilde{k}$.
This linear term can simply be removed by the following shift
\begin{equation}\label{shift}
\theta^{s-}\to\theta^{s-}+\frac{k_Fx}{u^{s-}K^{s-}}\frac{\sqrt{2(\alpha^2+\beta^2)}}{\pi}
\end{equation}
and similarly for $\theta^{s+}$ so that $H_q+H_{\textrm{SO}}\to H_q +\mbox{const}$. However, this shift has to
be carefully taken into account for the non-quadratic interaction
terms. As we will see below, it will induce slow oscillations in space
for some of these terms.

\subsection{Bosonized interactions}
In addition to the quadratic bosonic Hamiltonian we have the set of
backscattering, inter-band scattering, and double spin-flip scattering interactions,
Eqs.~\eqref{int_bs}-\eqref{int_dsf}. The vertex operator,
Eq.~\eqref{vertex}, allows a straightforward bosonization of these
interactions. Firstly, the $g_1$ backscattering interactions become
 \begin{eqnarray}
\label{back_bos}
H_{1 \perp} &=& \frac{g_{1 \perp}}{\left(\pi \alpha \right)^2} \int \ud x \cos\left[\sqrt{4\pi} \phi^{s +} \right]\cos\left[\sqrt{4\pi} \phi^{s -} \right]\,, \nonumber \\
H'_{1 \parallel} &=& -\frac{g'_{1 \parallel} }{\left(\pi \alpha \right)^2}\int \ud x \cos\left[\sqrt{4\pi} \theta^{s -} \right]\cos\left[ \sqrt{4\pi} \theta^{c -} \right]\,,\nonumber \\
H_{1 \perp}'  &=& \frac{g_{1 \perp}'}{\left(\pi \alpha \right)^2}  \int \ud x \cos \left[\sqrt{4\pi} \phi^{s +} \right]\cos\left[ \sqrt{4\pi} \theta^{c -} \right]\,,\,\qquad\\
\bar{H}_{1 \perp}' &=& \frac{\bar{g}_{1 \perp}'}{\left(\pi \alpha \right)^2} \int \ud x \cos \left[ \sqrt{4\pi} \phi^{s +} \right]\cos \left[ \sqrt{4\pi} \theta^{s -} \right]\,.\nonumber
 \end{eqnarray}
Secondly, there is a $g_2$ processes
 \begin{equation}
   H_{2 \perp}' = \frac{g_{2 \perp}'}{\left(\pi \alpha \right)^2} \int \ud x \cos \left[ \sqrt{4\pi} \phi^{s -} \right]\cos \left[ \sqrt{4\pi} \theta^{c -} \right]\,.
 \end{equation}
Lastly, there is a $g_4$ process
\begin{equation}
\bar{H}_{4 \perp}' = \frac{\bar{g}_{4 \perp}'}{\left(\pi \alpha \right)^2} \int \ud x \cos \left[ \sqrt{4\pi} \phi^{s -} \right]\cos \left[ \sqrt{4\pi} \theta^{s -} \right]\,.
 \end{equation}
Several of these terms are shown schematically in Fig.~\ref{scattering}.

The allowed double spin-flip backscattering interactions, Eq.~\eqref{int_dsf0}, in bosonized form are given by
 \begin{eqnarray}
H_{1s} &=& \frac{g_{1s}}{\left(\pi \alpha \right)^2} \int \ud x \cos\left[\sqrt{4\pi} \theta^{s +} \right]\cos\left[\sqrt{4\pi} \theta^{s -} \right]\,,\nonumber\\
H'_{1s} &=& \frac{g'_{1s} }{\left(\pi \alpha \right)^2}\int \ud x \cos\left[\sqrt{4\pi} \theta^{s+} \right]\cos\left[ \sqrt{4\pi} \theta^{c -} \right]\,,\qquad\\
\bar{H}'_{1s} &=& \frac{\bar{g}'_{1s}}{\left(\pi \alpha \right)^2}  \int \ud x \cos \left[\sqrt{4\pi} \phi^{s-} \right]\cos\left[ \sqrt{4\pi} \theta^{s+} \right].\nonumber
 \end{eqnarray}
 The last two spin-flip interactions which
 contribute, Eq.~\eqref{int_dsf}, are
\begin{equation}
\bar{H}_{4s} = \frac{\bar{g}_{4s}}{\left(\pi \alpha \right)^2} \int \textup{d}x \cos \left[ \sqrt{4\pi} \phi^{s +} \right]\cos \left[ \sqrt{4\pi} \theta^{s +} \right]\,,
 \end{equation}
and
\begin{eqnarray}
H_{4s} &=& \frac{g_{4s}}{\left(\pi \alpha \right)^2} \int \ud x \left\{\prod_\beta\cos \left[ \sqrt{4\pi} \theta^{s \beta} \right]\cos \left[ \sqrt{4\pi} \phi^{s \beta} \right]\right.\nonumber\\
&&\hspace{0.8cm}+\left.\prod_\beta\sin \left[ \sqrt{4\pi} \theta^{c\beta} \right]\sin \left[ \sqrt{4\pi} \phi^{c\beta} \right]\right\}\,.
 \end{eqnarray}

If we now shift the $\theta^{s-}$ and $\theta^{s+}$ fields to compensate for the spin
 splitting in the bands, see Eq.~\eqref{shift}, then the
 backscattering terms $H'_{1\parallel},\,\bar H'_{1\perp}$ and $\bar
 H'_{4\perp}$ and all the double spin-flip processes become slowly oscillating. These oscillations will suppress
 the interactions at the lowest temperatures, when the correlation
 length $\xi\sim v_F/T\gg1/k^b_{\textrm{SO}}$. They will, however, still be
 present in the RG flow at intermediate temperatures,
 $\varepsilon^b_{\textrm{SO}}\ll T\ll \varepsilon_F$.  In section \ref{sec_phase}
 we will consider the phase diagram at zero temperature as well as an
 effective phase diagram in the intermediate temperature regime.

\section{$SU (2)$ symmetry and spin density correlation functions}
\label{sec_symmetry}
Before we continue with the calculation of the phase diagram in the
presence of the interaction terms, we want to study first the spin-spin correlation functions
for the quadratic Hamiltonian \eqref{H_quad}. In particular, we want
to find out what conditions are imposed on the parameters of the
two-band model at the point where $SU(2)$ spin symmetry is restored.
Contrary to the usual single band model where this leads to $K_s=1$ it
is not \emph{a priori} clear if a similar condition also holds in the
two-band case.

For the following calculations it is convenient to express the new
fields $\phi^{\nu\beta}$ and $\theta^{\nu\beta}$, introduced to
diagonalize the quadratic part of the Hamiltonian, in terms of new
chiral fields determined by
\begin{equation}
 \phi^{\nu \beta}_\delta = \frac{1}{\sqrt{2}} \left( \frac{\phi^{\nu \beta}}{\sqrt{K^{\nu\beta}}} - \delta \,\sqrt{K^{\nu\beta}}\theta^{\nu \beta} \right)\,,
\end{equation}
where $\delta = \pm$ is once again a direction index. These new fields describe the chiral excitations of the system moving either to the left, $\delta=-$, or to the right, $\delta=+$.
In this basis the appropriate time ordered correlation functions of the bosonic fields are given by
\begin{eqnarray}\label{chiralcorrelation}
G_{\nu\beta\delta}(x,t)&\equiv& \left\langle T_t \left( \phi_\delta^{\nu \beta} (x,t) - \phi_\delta^{\nu \beta} (0) \right)^2 \right\rangle\\\nonumber
&=& \frac{1}{\pi} \ln \left[\frac{\alpha + \sgn(t)\im \left( u^{\nu \beta} t - \delta x \right)}{\alpha} \right]\,.
\end{eqnarray}
Using this correlation function one can calculate the correlations of
the oscillating parts of the spin density waves
\begin{eqnarray}
 S_{SDW}^j (x,t) = \sum_{\substack{\sigma,\sigma';\\(r,b)\neq(r',b')}} \textup{e}^{\im \left( r' \eta_{b'} k_{F b'} - r \eta_{b} k_{F b} \right) x} \\\nonumber
 \times \psi_{\sigma r b}^\dagger (x,t) \sigma_{\sigma \sigma'}^j \psi_{\sigma' r' b'} (x,t)
\end{eqnarray}
where $\sigma^j$ are the Pauli matrices for $j = x,y,z$.

For the oscillating parts of the spin density wave correlation function in the $x$ or $z$ direction (in the $y$ direction it is trivially equivalent to that in $x$) we obtain
\begin{eqnarray}\label{sdw}
 \left\langle S_{SDW}^{x,z}(x,0) S_{SDW}^{x,z} (0,0) \right\rangle =
\frac{1}{\left( \pi \alpha \right)^2}   \left[\sum_b
\frac{\cos[2k_{Fb}x]}{|x/\alpha|^{\varepsilon_{c}+\varepsilon_{s}^{x,z}}}\right.\nonumber\\
+\frac{\cos[(k_{F1}+k_{F2})x]}{|x/\alpha|^{\bar{\varepsilon}_{c}+\bar{\varepsilon}_{s}^{x,z}}}
\left.+\frac{\cos[(k_{F1}-k_{F2})x]}{|x/\alpha|^{\varepsilon_{cf}+\varepsilon_{sf}^{x,z}}}\right]\,.\qquad
\end{eqnarray}
There are three different types of competing density waves present.
The first is backscattering which preserves the band index, with exponents $\varepsilon_\nu^{x,z}$, the direct
analogue of backscattering in a single band system. The second is
backscattering which mixes the bands, $\bar{\varepsilon}_\nu^{x,z}$. Finally there is a forward
scattering process which scatters between the bands, $\varepsilon_{\nu
  f}^{x,z}$. The exponents are summarized in Table
\ref{table_exponents}.
\begin{table}
\begin{ruledtabular}
{ \renewcommand{\arraystretch}{1.5}
 \renewcommand{\tabcolsep}{0.2cm}
\begin{tabular}{c|c}
Charge exponent & Spin exponent \\\hline
$\varepsilon_{c}=\frac{1}{2}\left(K^{c+}+K^{c-}\right)$ & $\varepsilon_{s}^{z}=\frac{1}{2}\left(K^{s+}+K^{s-}\right)$ \\
 & $\varepsilon_{s}^{x}=\frac{1}{2}\left(\frac{1}{K^{s+}}+\frac{1}{K^{s-}}\right)$\\\hline
$\bar{\varepsilon}_{c}=\frac{1}{2}\left(K^{c+}+\frac{1}{K^{c-}}\right)$ & $\bar{\varepsilon}_s^{z}=\frac{1}{2}\left(K^{s+}+\frac{1}{K^{s-}}\right)$ \\
 & $\bar{\varepsilon}_s^{x}=\frac{1}{2}\left(K^{s-}+\frac{1}{K^{s+}}\right)$ \\\hline
$\varepsilon_{cf}=\frac{1}{2}\left(K^{c-}+\frac{1}{K^{c-}}\right)$ & $\varepsilon_{sf}^{z}=\frac{1}{2}\left(K^{s-}+\frac{1}{K^{s-}}\right)$ \\
 & $\varepsilon_{sf}^{x}=\frac{1}{2}\left(K^{s+}+\frac{1}{K^{s+}}\right)$ \\
\end{tabular}}
  \caption{\label{table_exponents} The exponents for different scattering processes in the spin density wave correlations, Eq.~\eqref{sdw}.}
\end{ruledtabular}
\end{table}

For $SU (2)$ symmetry to hold the spin density-spin density correlation
should be the same with respect to any spatial direction. This is clearly
always fulfilled for the charge exponents, but gives us a set of conditions
for the spin exponents. In the case of equivalent bands (the more general
case is explained in Appendix \ref{appendix_rotations}) this imposes
\begin{equation}
 K^{s +} = K^{s -} = 1\,,
\end{equation}
which is indeed in direct analogy to the usual single band Luttinger
liquid condition.

\section{Phase Diagram}\label{sec_phase}
To see how the backscattering terms, Eq.~\eqref{int_bs}, and the double
spin-flip terms, Eqs.~(\ref{int_dsf0}, \ref{int_dsf}), change the behaviour
of the system we perform a first order RG analysis on them. Note that
away from the $SU (2)$ symmetric point the terms become unambiguously
irrelevant or relevant forgoing the need for a more complicated second
order treatment.

The standard first order renormalization group analysis yields for the
interactions of the system a set of independent equations for the flow
of the coupling constants $g_i$, $\bar{g}_i$, and $\bar{g}'_i$ under a
change of the length scale $l$
 \begin{equation}
  \frac{1}{g_{i}} \frac{\textup{d} g_{i}}{\textup{d} l} = 2 - \gamma_i\,.
 \end{equation}
 The $\{\gamma_i\}$ are then the scaling dimensions of the
 corresponding scattering terms.  These scaling dimensions can be
 easily extracted by power counting. Several of the interaction terms
 are always irrelevant, or at best marginal in an $SU(2)$ symmetric
 system, and we first list these together here:
 \begin{eqnarray}
\bar{\gamma}'_{4 \perp}&=&K^{s -} +\frac{1}{K^{s -}}\,,\quad \bar{\gamma}_{4s}=K^{s +}+\frac{1}{K^{s +}}\,,\nonumber\\
\gamma_{4s}&=&K^{s+}+\frac{1}{K^{s +}}+K^{s-}+\frac{1}{K^{s -}}\, .
 \end{eqnarray}
The remaining backscattering scaling dimensions are
 \begin{eqnarray}
\gamma_{1 \perp}&=&K^{s +}+K^{s -}\,,\nonumber\\
\gamma'_{1 \parallel}&=&\frac{1}{K^{s -}}+\frac{1}{K^{c -}}\,,\quad \gamma'_{1 \perp}=K^{s +}+\frac{1}{K^{c -}}\,, \\
\bar{\gamma}'_{1 \perp}&=&K^{s +}+\frac{1}{K^{s -}}\,,\quad\gamma'_{2 \perp}=K^{s -}+\frac{1}{K^{c -}}\,.\nonumber
 \end{eqnarray}
Similarly, we find for the double spin-flip scattering terms
 \begin{eqnarray}
\gamma_{1s}&=&\frac{1}{K^{s+}}+\frac{1}{K^{s -}}\,,\quad\gamma'_{1s}=\frac{1}{K^{s+}}+\frac{1}{K^{c -}}\,,\nonumber\\
\bar{\gamma}'_{1s}&=&K^{s -}+\frac{1}{K^{s +}}\,.
 \end{eqnarray}
 Finally, the two possibly relevant single spin-flip interactions
 which could modify the intermediate phase diagram, see
 Eq.~\eqref{relsf}, have scaling dimensions
\begin{eqnarray}
\gamma'_{1sf} &=& \frac{1}{K^{c-}}+\frac{1}{4}\left(K^{s+}+\frac{1}{K^{s+}}+K^{s-}+\frac{1}{K^{s-}}\right),\nonumber\\
\label{spin-flip_1}
\gamma'_{2sf} &=& K^{c-}+\frac{1}{4}\left(K^{s+}+\frac{1}{K^{s+}}+K^{s-}+\frac{1}{K^{s-}}\right).\qquad
\end{eqnarray}
All other single spin-flip processes have a scaling dimension
$\gamma_{sf}>2$ and are thus irrelevant.

For a relevant interaction term the coupling constant
grows whilst lowering the temperature. Thus the bosonic
fields present in this interaction will get pinned to the values which
minimize the energy, leading to a gap in the corresponding dual mode.
We will follow the standard notation where C$x$S$y$ is a phase of the
system with $x$ gapless charge and $y$ gapless spin
modes.\cite{Balents1996} Note, however, that of course only one of the
dual fields, $\phi^{\nu\beta}$ and $\theta^{\nu\beta}$, can be pinned since the
commutation relation between them must be preserved.\cite{Chang2007}

\begin{figure}
\includegraphics[width=0.4\textwidth]{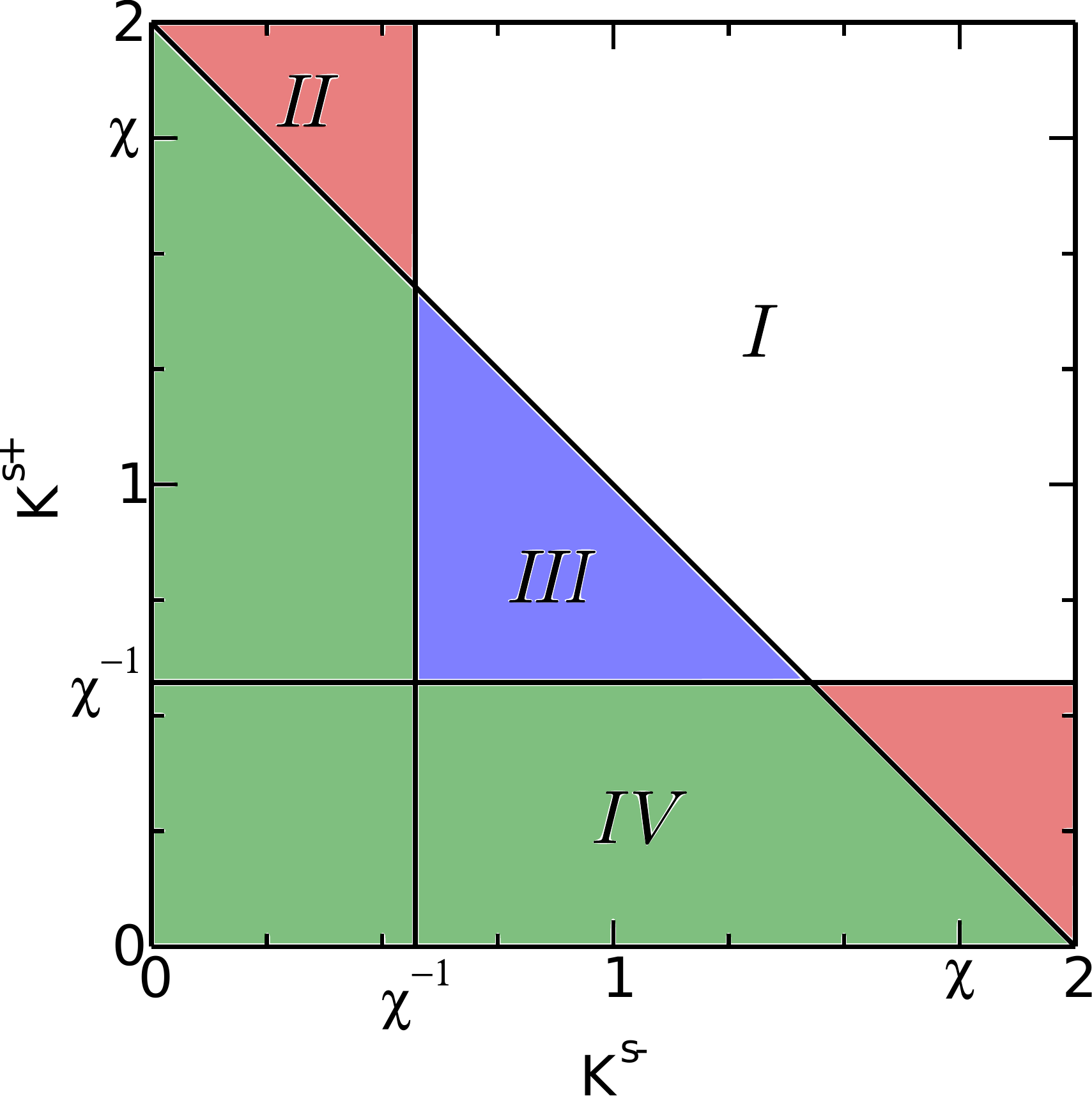}
\caption[Phases]{The phase diagram at temperatures $T\ll
  \varepsilon_{\textrm{SO}}$ with $\chi$ as defined in \eqref{chi}. Here we
  have used $K^{c-}=0.7$.  The solid lines are the separatrices between
different phases and the C2S2 Luttinger liquid phase is the unshaded white
region. The red regions marked (II) are C1S1 phases, blue (III) are C2S0 phases, and green (IV) are C1S0 phases.}
\label{Phase_Plot_0T}
\end{figure}

As we are interested in models without $SU(2)$-symmetry it is most
convenient to plot the phase diagram for $K^{s\pm}$ with $K^{c-}$ a
parameter. To establish the phase diagram we define for convenience
\begin{equation}
\label{chi}
 \chi = \frac{K^{c -}}{2 K^{c -} - 1}\,.
\end{equation}
The large number of scattering terms for the two-band model leads to a very rich phase diagram,
which contains a Luttinger liquid region; see Fig.~\ref{Phase_Plot_0T}. This region
becomes enlarged for strong interactions as the horizontal and
vertical separatrices are $K^{c-}$ dependent and for $K^{c-}\to 0.5$
$\chi\to\infty$. I.e.~these separatrices are completely removed from
the phase diagram for $K^{c-}\leq 0.5$. Conversely, as $K^{c-}\to 1$ we
  have $\chi\to 1$, reducing the extent of the C2S2 phase.

\begin{figure}
\includegraphics[width=0.4\textwidth]{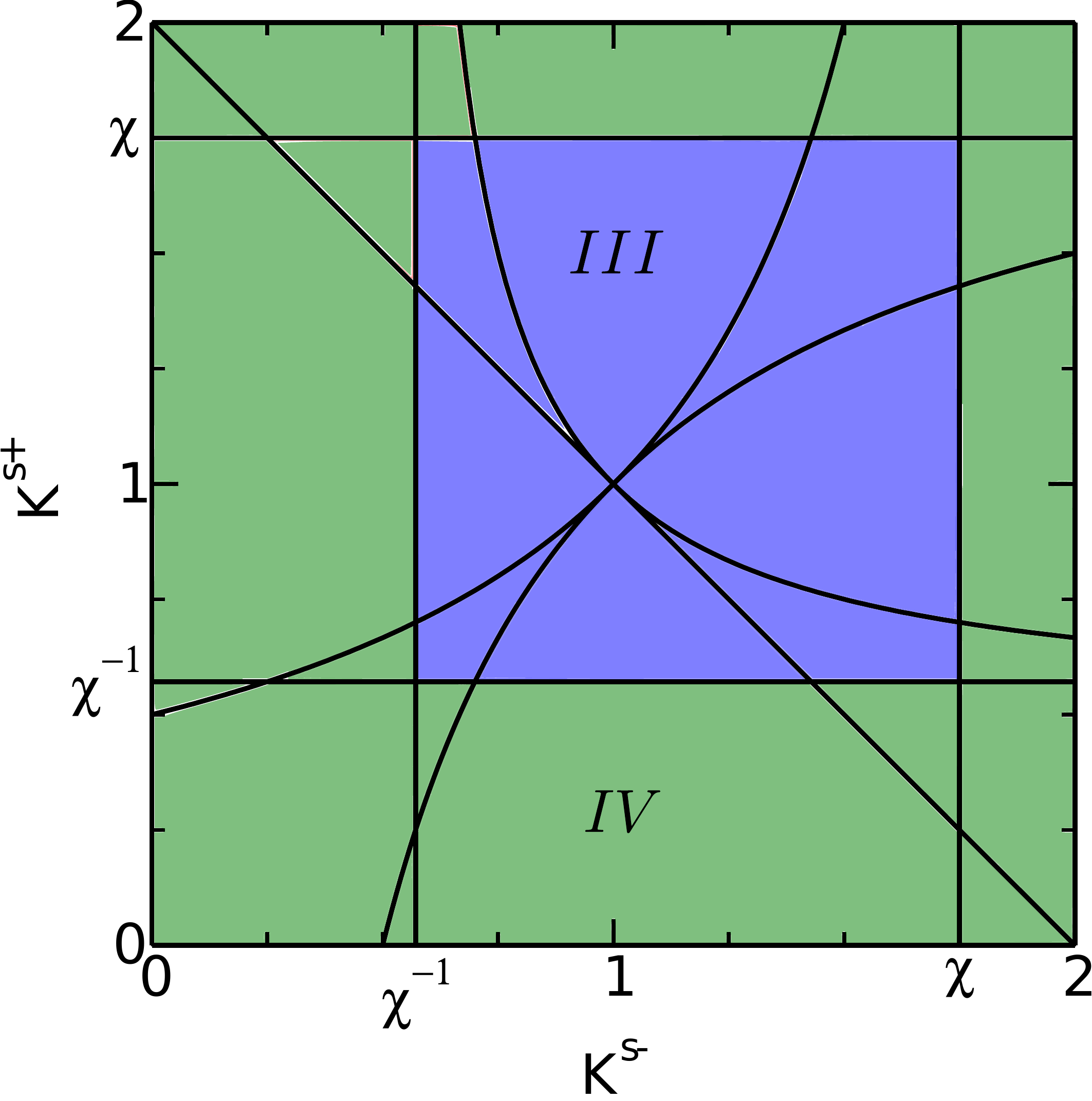}
\caption[Phases]{(Color online) The
  effective phase diagram at $\varepsilon^b_{SO}\ll T\ll
  \varepsilon_F$. At these temperatures the slowly oscillating
  interaction terms have to be kept and lead to additional phases
  where modes appear to be thermally activated. As in
  Fig. 4 we exemplarily show the diagram for
  $K^{c-}=0.7$.  The solid lines are the separatrices between different
  phases. The blue regions marked (III) are C2S0 phases, and green
  (IV) are C1S0 phases.}
\label{Phase_Plot_Int_T}
\end{figure}

However, this phase diagram will only hold at the lowest temperatures
where the slowly oscillating scattering terms containing the $\theta^{s-}$ and
  $\theta^{s+}$ fields can be neglected. At temperatures
$\varepsilon^b_{\textrm{SO}}\ll T\ll \varepsilon_F$ these scattering terms also
have to be kept \cite{GarateAffleck} leading to additional sections of
the phase diagram where some modes will appear thermally activated and
the corresponding spectral weight strongly suppressed. A well known
example for such behavior is the one band Hubbard model solvable by
Bethe ansatz. At filling $n=1$ and on-site interaction $U>0$ the
charge mode is gapped (Mott insulator) while there is no gap away from
half filling. However, at $n=1\pm\epsilon$ with $|\epsilon|\ll 1$ the
cosine scattering term responsible for the Mott transition will
oscillate only very slowly. As a consequence, the charge
compressibility will look thermally activated as in the half filled
case at high and intermediate temperatures with a steep increase
visible only at the lowest temperatures. \cite{JuttnerKlumper} The
effective phase diagram including the phases which appear to be gapped
at temperatures $\varepsilon^b_{\textrm{SO}}\ll T\ll \varepsilon_F$ is shown in
Fig.~\ref{Phase_Plot_Int_T}.

In the experiments considering self-organized gold chains on a Ge(001)
surface a Luttinger liquid (C2S2) phase appears to be
seen.\cite{Blumenstein2011} Given the large number of scattering terms
for a system with four Fermi points this may seem surprising, although our
results show that it is possible if the Luttinger liquid
parameters are in the right range. However, these restrictions on the Luttinger parameters also mean that the structure
of the spin density waves in the Luttinger liquid are constrained. In
particular, we are interested in what these constraints mean for the decay
of the in plane and perpendicular components of the spin correlation
function. Physically, we might expect that the spins are lying mainly
within the surface with the perpendicular component being
comparatively smaller. This can be used as a consistency check to see
if the formation of a Luttinger liquid in this surface system is
reasonable.

If we assume that the system is in the low temperature C2S2 phase,
shown in Fig.~\ref{Phase_Plot_0T}, then we have a set of constraints
on the Luttinger parameters. The separatrices of this phase are
composed of the $\gamma'_{1\perp}$,
$\gamma'_{2\perp}$ and $\gamma_{1\perp}$ interaction processes. There are two constraints which
involve the charge Luttinger parameter:
\begin{equation}
 K^{s+}>\chi^{-1}\quad \text{and} \quad K^{s-}>\chi^{-1}\,.
\end{equation}
Additionally we require the following to hold between the spin Luttinger parameters:
\begin{equation}
 K^{s+} + K^{s-} >2\,.
\end{equation}
From these general considerations some
conclusions follow about the spin density wave correlation functions
in the Luttinger liquid phase, see Eq.~\eqref{sdw}. If $K^{s+}K^{s-}>1$---which is true
  for most of the C2S2 phase---then we find for the intra band
  exponents $1\leq \varepsilon^z_s$  and $\varepsilon^x_s<\varepsilon^z_s$. I.e. the
spin-density spin-density correlation function decays quicker in the
out of plane direction, as one may expect. For interband
  backscattering, $\bar\varepsilon_s^{x,z}$, neither in plane nor out
  of plane correlations are necessarily preferred.
For forward scattering both
$\varepsilon^{x,z}_{sf}\geq1$, but the relation between them is not
fixed. In general we can say nothing for the relative in and out of
plane power laws of the inter band forward scattering terms. The
actual spin order in the system will be the result of the competition
between these possible processes.

\subsection{Characterization of phases with gapped modes}
If the system is in a phase where at least one mode is gapped, single
particle correlations will in general no longer be described by power
laws but will decay exponentially.\cite{Chang2007} In particular, the
spectral function will always show exponential decay in phases with
gapped modes. However, if we consider many-particle correlations which
do not involve the gapped modes then they will still behave as power
laws. It is therefore standard practice to characterize a gapped phase
by the many-particle correlation function which shows the slowest
decay.

To consider a concrete example, let us assume that the $g_{1\perp}$
backscattering term, Eq.~\eqref{back_bos}, is relevant. Then the
fields $\phi^{s+}$ and $\phi^{s-}$ are pinned to the values which
minimize the energy of this backscattering term. Every correlation
function which involves at least one of the dual fields $\theta^{s+}$
or $\theta^{s-}$ is then exponentially decaying. The many-body
correlation functions $\langle O(x,t)O(0,0)\rangle$ which will still
show a power law decay are those with an operator $O$ which does not
contain these two dual fields. All operators for which this is true
can be determined by using the bosonization dictionary given in
Appendix \ref{appendix_bosons}. For the considered example we find that the two singlet pair
operators $O_{S,\sigma;br,br}=\Psi_{\sigma br}\Psi_{\bar\sigma br}$
and $O_{S,\sigma;br,b\bar r}=\Psi_{\sigma br}\Psi_{\bar\sigma b\bar
  r}$ as well as the charge density wave $O_{CDW,\sigma;br,b\bar
  r}=\Psi^\dagger_{\sigma br}\Psi_{\sigma b\bar r}$ will be the pair
correlation functions which decay with a power law. Which one of these
shows the slowest decay depends on the values of the Luttinger
parameters.

For all other phases with two gapped modes we can use a similar
procedure.  In addition to the already introduced singlet pair
operator $O_S$ and charge density wave operator $O_{CDW}$ also the
triplet pair correlations with $O_{T,\sigma; r b, r' b'} =
\psi_{\sigma r b} \psi_{\sigma r' b'}$ and the spin density wave
$O_{SDW,\sigma;br,b'r'}=\psi_{\sigma r b}^\dagger \psi_{\bar{\sigma} r' b'}$ can show
power law decay. A list of phases with the corresponding pinned fields
and the gapless many-body correlations is given in Table
\ref{table_twoparticle}.
\begin{table}
 \centering
 \begin{tabular}{c|c|c|c}
  Phase &Pinned Fields &Pair Operator &Density Waves \\\hline
  C2S0 & $\phi^{s -}$, $\phi^{s +}$ & $O_{S, \sigma; r b, r b}$, $O_{S, \sigma; r b, \bar{r} b}$ & $O_{CDW, \sigma; r b, \bar{r} b}$ \\
  C2S0 & $\phi^{s -}$, $\theta^{s +}$ & $O_{S, \sigma; r b, r b}$  &  $O_{SDW, \sigma; r b, \bar{r} \bar{b}}$ \\
  C1S1 & $\theta^{c -}$, $\phi^{s -}$ & $O_{T, \sigma; r b, r \bar{b}}$ & $O_{SDW, \sigma; r b, \bar{r} \bar{b}}$ \\
  C1S1  &$\theta^{c -}$, $\theta^{s +}$ & $O_{S, \sigma; r b, r \bar b}$, $O_{T, \sigma; r b, \bar r b}$    & $O_{SDW, \sigma; r b, \bar{r} \bar{b}}$
 \end{tabular}
 \caption{Pair operators and density waves which lead to algebraically decaying correlations
in phases with two gapped modes. All possible pairs of gapped modes for the zero temperature
phase diagram are included.}
 \label{table_twoparticle}
\end{table}
For the phases with three gapped modes, four-particle correlators can
be constructed using the bosonization dictionary which do not contain
the fields dual to the pinned fields. However, these correlations are
of little physical use and we do not give them here explicitly.

\section{Spectral function and density of states}\label{sec_spectral}
For the system of gold wires on a Ge(001) surface the density of
states (DOS) has been measured experimentally and found to show power
law scaling. As discussed in the previous section, such a power law
scaling will only occur if all modes are gapless, i.e. the system is in the C2S2
phase.

In this section we will calculate the spectral function $A(q,\omega)$
in the C2S2 phase from which the density of states $\nu(\omega)$ can
be obtained by a momentum integration. The spectral
function itself may be measurable in a suitable experiment by ARPES. While the
DOS gives only information about a certain combination of the
Luttinger liquid parameters and thus by itself does not allow a check
if the predictions of our model for the extent of the C2S2 phase are
consistent with experiment, the spectral function will, in principle,
allow one to determine all four Luttinger parameters separately and thus allow a
full consistency check.

The spectral function for the interacting fermionic model can be
calculated directly using the bosonic Luttinger liquid
representation.\cite{Meden1992,Voit1995} In general, it is defined as $A(k,\omega)=-\frac{1}{\pi}\text{Im}\,
G^{\textrm{ret}}(k,\omega)$ where $G^{\textrm{ret}}(k,\omega)$ is the
retarded Green's function. Equivalently we can write, with $\br=(x,t)$,
\begin{equation}
A(k,\omega)=\frac{1}{2\pi}\sum_\sigma\int\!\!\ud x\ud t \e^{\im(\omega t-kx)}\left\langle\left\{\psi_{\sigma}(\br),\psi^\dagger_{\sigma}(0)\right\}\right\rangle\,.
\end{equation}
After linearization we are therefore interested in
\begin{equation}
\im G^>_{\sigma}(\br)=\sum_{br}\e^{\im r\eta_bk_{Fb}x}\langle\psi_{\sigma br}(\br)\psi^\dagger_{\sigma br}(0)\rangle\,,
\end{equation}
and a similar term for $\im G^<_{\sigma}(\br)$ where the two fermionic
operators are interchanged.
Ignoring the small spin-orbit splitting we can use the bosonic
  Green's function, Eq.~\eqref{chiralcorrelation}, and find
\begin{equation}\label{greensfunction}
\im G^>_{\sigma}(\br)=\frac{1}{2\pi \alpha}\sum_{br}\e^{\im r\eta_bk_{Fb}x}\prod_{\nu\beta\delta}
\e^{-\pi\xir G_{\nu\beta\delta}(\br)}
\end{equation}
where $\delta=\pm$ denotes the chiral component of the field and
\begin{equation}
 \xir=\frac{1}{16}\left[\sqrt{K^{\nu\beta}}-\frac{\delta r}{\sqrt{K^{\nu\beta}}}\right]^2\,,
\end{equation}
which arises from the rotation between the original $\phi_{\sigma br}$ field and the chiral mode $\phi_{\delta}^{\nu\beta}$.
In the space and time representation this leads to
\begin{eqnarray}\label{spectralfunction2}
A(\br)&\sim&\frac{1}{2\pi^2 \alpha}\sum_{\sigma br}\e^{\im r\eta_bk_{Fb}x}
\prod_{\nu\beta\delta}\left[\frac{\alpha}{|x-\delta u^{\nu\beta}t|}\right]^{\xir}\quad
\end{eqnarray}
with
\begin{equation}\label{spectralfunction}
A(k,\omega)=\int\ud x\ud t\, \e^{\im\omega t-\im kx}A(\br)\,.
\end{equation}
This integral cannot be calculated analytically in full but we can
obtain the singular contributions\cite{Meden1992,Voit1995} which occur at $\omega = \pm u^{\nu \beta} |k|$.

Using Eqs.~\eqref{spectralfunction2} and \eqref{spectralfunction} the full spectral function can be decomposed into the sum
\begin{equation}\label{spec0}
A(k,\omega)=\sum_{rb}A_r(k-r\eta_bk_{Fb},\omega)\,.
\end{equation}
The spectral function has the symmetry $A_+(q,\omega)=A_-(-q,\omega)$, and therefore we focus only on $A_+(q,\omega)$.
To calculate these contributions we first order the four velocities in order of increasing magnitude such that $u^{\nu_1 \beta_1}<u^{\nu_2 \beta_2}<u^{\nu_3 \beta_3}<u^{\nu_4 \beta_4}$, where as before $\nu_i=c,s$ and $\beta_i=\pm$. Then we make a relabeling such that $u_i=u^{\nu_i \beta_i}$. Now for positive momenta $q>0$ we find in the vicinity of $\omega \approx u_1q$
\begin{eqnarray}\label{spec1}
 A_{r} (q, \omega)  &\sim& \Theta \left( \omega - u_1 q \right)\\\nonumber&& \left( \omega - u_1 q \right)^{\left(\gamma_1+2\gamma_2+2\gamma_3+2\gamma_4-(r+1)/4\right)/2}\,.
\end{eqnarray}
The exponents are given by
\begin{equation}
\gamma_i=\frac{1}{8}\left[K^{\nu_i \beta_i}+\frac{1}{K^{\nu_i \beta_i}}-2\right]\,.
\end{equation}
At the remaining singular points, $\omega \approx u_i q$ with $i\in\{2,3,4\}$, the spectral function behaves as
\begin{equation}\label{spec2}
 A_{r} (q, \omega)  \sim \left\lvert \omega - u_i q \right\rvert^{\left(\gamma_i+2\sum_{j\neq i}\gamma_j-(r+1)/4\right)/2}.
\end{equation}
The full $A_{r} (q>0, \omega)$ is the product over Eqs.~\eqref{spec1} and \eqref{spec2}.

\begin{figure}
\includegraphics*[width=1.0\columnwidth]{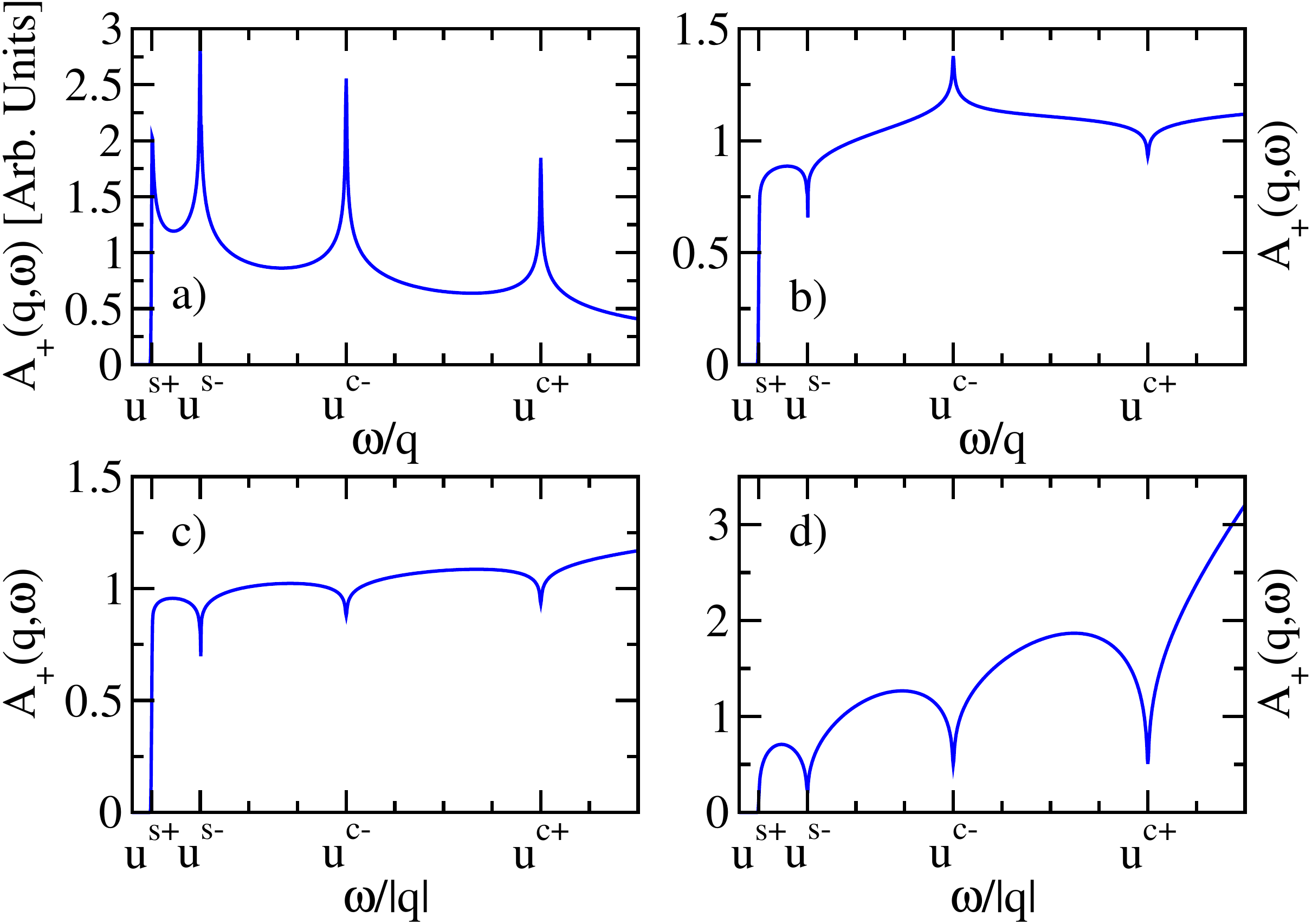}
\caption{(Color online) The spectral function,  Eq.~\eqref{spectralfunction}. Insets a) and c) are plotted for $K^{c+}=K^{c-}=0.7$, $K^{s+}=1.3$ and $K^{s-}=1$. Insets b) and d) are plotted for $K^{c+}=0.3$, $K^{c-}=0.5$, $K^{s+}=1.7$ and $K^{s-}=1.3$. The top row, a) and b), are for positive momenta, the bottom row, c) and d), are for negative momenta. The spectral function has been convoluted with a Gaussian resolution function of width $\omega/u^{c-}|q|=0.001$.}
\label{Spectrum}
\end{figure}

For negative momenta,
$q<0$, we find near $\omega \approx u_1|q|$
\begin{eqnarray}
 A_{r} (q, \omega)  &\sim& \Theta \left( \omega + u_1 q \right)\\\nonumber&& \left( \omega+ u_1 q \right)^{\left(\gamma_1+2\gamma_2+2\gamma_3+2\gamma_4+(r-1)/4\right)/2}\,.
\end{eqnarray}
At the remaining singular points, $\omega \approx u_i q$ with $i\in\{2,3,4\}$, the spectral function behaves as
\begin{equation}
 A_{r} (q, \omega)  \sim \left\lvert \omega+ u_i q \right\rvert^{\left(\gamma_i+2\sum_{j\neq i}\gamma_j+(r-1)/4\right)/2}.
\end{equation}
The spectral function for positive and negative momenta and $r=+$ is plotted in Fig.~\ref{Spectrum} for two different parameter sets. The slope of the divergences and cusps yields information about the four Luttinger parameters, which would allow a consistency check on our model. The existence of cusps vs divergences is dependent on the values of the Luttinger parameters, as can be seen from a comparison of Figs.~\ref{Spectrum}(a) and Figs.~\ref{Spectrum}(b). From Eq.~\eqref{spec0} one can see that the measured spectral function will consist of four sets of peaks and cusps around the four Fermi points. The two positive Fermi momenta will show the same structure, as will the negative Fermi points.

One of the classic signatures of a Luttinger liquid, taken as
indication of a Luttinger liquid state in the experiment of
Blumenstein et al.\cite{Blumenstein2011} on the monatomic gold
chains, is the power law suppression of the DOS near the
Fermi energy.\cite{Braunecker2012,Schuricht2013} The experimental result was analyzed with the single band DOS. Here we now derive the equivalent formula for the
appropriate two-band model directly from the spectral function
\begin{eqnarray}\label{dos}
\nu(\omega)&\sim&\sum_q A(q,\omega)=A(x=0,\omega)\\\nonumber
&\sim&\frac{1}{2\pi^2\alpha}\int\ud t\,\e^{\im \omega t}\sum_{\sigma br}\prod_{\nu\beta\delta}\left(\frac{\alpha}{\delta u^{\nu\beta}t}\right)^{\xir}\,.
\end{eqnarray}
We are here only interested in the power law suppression, which can be gained directly by power counting, and gives
$\nu(\omega)\sim\omega^{\gamma}$ with the exponent
\begin{equation}
\gamma=\sum_{\nu\beta\delta}\xir-1
=\frac{1}{8}\sum_{\substack{\nu=c,s\\\beta=\pm}}\left[K^{\nu \beta}+\frac{1}{K^{\nu \beta}}-2\right]\,.
\end{equation}
As a comparison, the standard results for this exponent for a spinless
as well as for a spinful single band Luttinger liquid are given in
Table \ref{dos_table}.
\begin{table}
\begin{ruledtabular}
{ \renewcommand{\arraystretch}{1.5}
 \renewcommand{\tabcolsep}{0.2cm}
\begin{tabular}{p{1.4cm}@{}|c|c}
Model & General & $SU(2)$ symmetric \\\hline
Spinless & $\frac{1}{2}\left[K+\frac{1}{K}-2\right]$ & N.a. \\\hline
Single band & $\frac{1}{4}\sum\limits_{\nu}\left[K_\nu+\frac{1}{K_\nu}-2\right]$ &  $\frac{1}{4}\left[K_c+\frac{1}{K_c}-2\right]$ \\\hline
Two-band & $\frac{1}{8}\sum\limits_{\nu\beta}\left[K^{\nu\beta}+\frac{1}{K^{\nu\beta}}-2\right]$ &  $\frac{1}{8}\sum\limits_{\beta}\left[K^{c\beta}+\frac{1}{K^{c\beta}}-2\right]$ \\
\end{tabular}}
  \caption{\label{dos_table} The exponents for power law suppression of the DOS near the Fermi energy: $\nu(\omega)\sim\omega^\gamma$. Listed are the exponents, $\gamma$, for a spinless, a standard single band, and a two-band Luttinger liquid. Also listed are the cases for $SU(2)$ symmetric models where $K^{s+}=K^{s-}=K_s=1$.}
\end{ruledtabular}
\end{table}

Near a boundary the DOS is suppressed with a different exponent. Exactly as in the single band case the bulk and boundary exponents are related by a conformal mapping.\cite{Giamarchi2004} In the two band Luttinger liquid, however, the relation between the bulk and the boundary exponent,
\begin{equation}
\gamma^{b}=\frac{1}{4}\sum_{\substack{\nu=c,s\\\beta=\pm}}\left[\frac{1}{K^{\nu \beta}}-1\right]\,,
\end{equation}
is no longer enough to independently check all four Luttinger parameters from DOS measurements alone.

Similarly the Green's function for finite temperatures can be calculated from the zero temperature case by a conformal mapping. The standard results for the finite temperature single band Luttinger liquid DOS then still apply with a suitably modified exponent.

\section{Discussion and conclusions}\label{conclusions}
Recent experiments on self-organized gold chains on a Ge(001) surface
have provided evidence for Luttinger liquid behavior. Other
experiments on similar surface systems have shown earlier that
spin-orbit coupling effects play an important role for the physics of
such systems with both Rashba and Dresselhaus-type couplings being
allowed by the reduced symmetry. Furthermore, the gold surface band is
found to cross the Fermi surface four times giving rise to two
separate electron pockets. The combination of spin-orbit coupling,
which breaks $SU(2)$ spin rotational symmetry, with the effective
two-band structure at low energies makes the gold chains very
different from other quasi one-dimensional systems as, for example,
carbon nanotubes or semiconducting nanowires where Luttinger liquid
behavior has also been seen.

In our paper we considered the low-energy effective theory of a
generic two-band model with spin-orbit coupling using bosonization. In
order to simplify the discussion we made a number of approximations.
First, we assumed that the two bands have equal Fermi velocities,
$v_{Fb}=v_{F\bar b}$. This seems to be approximately the case in the
experiments on gold chains. The generalization to the case of unequal
velocities is straightforward and is discussed in detail in
Appendix \ref{appendix_rotations}. Having unequal velocities for the two bands is a marginal
perturbation when starting from the symmetric band case. It can
therefore only affect the phase diagram of the model at second or
higher order in the RG flow which is beyond the scope of this paper.
Experimentally, this also means that such effects would only become
visible at very low temperatures if the considered system would remain
ideally one-dimensional.

Second, we ignored that the spin-orbit coupling leads to an effective
mixing of higher transverse modes into the lowest one which can make
the velocities of the spin-split band unequal, $v_{Fb\sigma}\neq
v_{Fb\bar\sigma}$.  This effect is the smaller the better the
confinement in the transverse direction is. For the gold chains the
dispersion perpendicular to the chain direction is flat so that
assuming a very strong confinement seems to be a good approximation.
Allowing for $v_{Fb\sigma}\neq v_{Fb\bar\sigma}$ activates additional
single spin-flip scattering processes which are otherwise forbidden by
symmetry. These scattering processes are, however, always oscillating
and therefore cannot modify the zero temperature phase diagram.

Using these two approximations we bosonized the two-band model with
spin-orbit coupling including all scattering terms which are allowed
by time reversal symmetry. A diagonalization of the kinetic part, the
density-density type interactions, and the spin-orbit coupling was
achieved by three separate rotations followed by a shift in one of the
bosonic fields. The remaining backscattering and double spin-flip
terms were then treated using a first order RG. This turns out to be
sufficient away from the $SU(2)$ symmetric point where most scattering
terms are either relevant or irrelevant. Some of the scattering terms
are very slowly oscillating in space due to the spin-orbit induced
splitting of the band and can thus be ignored at the lowest
temperatures. For this case we did calculate the full phase diagram
which turns out to consist of a Luttinger liquid (C2S2) phase as well
as phases where two or three out of the four modes (two spin and two
charge) are gapped. At intermediate temperatures $\varepsilon^b_{\textrm{SO}}\ll
T \ll\varepsilon_F$ the slowly oscillating scattering terms have to be kept in the RG flow. As a
  consequence, the spin modes might appear to be thermally activated
  even if all four modes are gapless at $T=0$.

For the experiment on gold chains this means that although the reduced
spin symmetry and the two bands allow for a large number of scattering
processes absent in the $SU(2)$ symmetric single band case, a
Luttinger liquid phase is still present in the $T=0$ phase diagram. This was
by no means a priori clear and shows that such a system could indeed
be a useful test bed for Luttinger liquid physics. Furthermore, our
theoretical study makes a clear prediction about the behavior of the
density of states and, more importantly, the full spectral function in
the C2S2 phase. While the derived formula for the density of states
shows that the exponent of the power law scaling in frequency does not
depend on a single Luttinger parameter as assumed in experiment but
rather on the four Luttinger parameters $K^{c\pm}$, $K^{s\pm}$, this
by itself does not allow one to verify if the predictions of the Luttinger
model are consistent with experiment. Here, additional measurements
would be desirable. First, the extent of the Luttinger liquid regime
in the phase diagram is restricted limiting the possible values
for the Luttinger parameters. From the limits on the Luttinger
parameters $K^{s\pm}$ we can, in particular, infer the decay of the
spin-spin correlations which potentially can be tested in experiment
by spin resolved STS. Probably even more promising is the measurement
of the full spectral function by ARPES which, in principle, allows for
the determination of all four Luttinger parameters separately and
therefore for a full consistency check with the appropriate Luttinger
model treated in this paper.

\begin{acknowledgments}
  The authors thank I.~Affleck and A.~Schulz for
  valuable discussions. We also acknowledge support by the DFG via the SFB/TR 49
  and by the graduate school of excellence MAINZ.
\end{acknowledgments}

\appendix

\section{Bosonization dictionary}\label{appendix_bosons}

Two useful formulae for bosonization are the vertex operator
\begin{equation}
 \psi_{\sigma br} (x) \sim \frac{1}{\sqrt{2 \pi \alpha}} \textup{e}^{\im  r \sqrt{2 \pi} \phi_{\sigma br} (x)}
\end{equation}
and the densities
\begin{equation}
\rho_{\sigma r b}(x)=  \psi_{\sigma br}^\dagger (x)\psi_{\sigma br} (x)=-\frac{1}{\sqrt{2 \pi}} \partial_x\phi_{\sigma br} (x)\,.
\end{equation}
The relation between these bosonic fields and the bosonic fields describing the diagonal normal modes is
\begin{eqnarray}
\phi_{\sigma br}&=&\frac{1}{\sqrt{8}}\left[\phi^{c+}+(-1)^b\phi^{c-}+\sigma(\phi^{s+}+(-1)^b\phi^{s-}) \right.\nonumber\\&&
\left.-r\left(\theta^{c+}+(-1)^b\theta^{c-}+\sigma(\theta^{s+}+(-1)^b\theta^{s-})\right)\right]\,.\nonumber\\&&
\end{eqnarray}
The chiral fields ($\delta=\pm$) are in turn determined by
\begin{equation}
 \phi^{\nu \beta}_\delta = \frac{1}{\sqrt{2}} \left( \frac{\phi^{\nu \beta}}{\sqrt{K^{\nu\beta}}} - \delta \,\sqrt{K^{\nu\beta}}\theta^{\nu \beta} \right)\,.
\end{equation}

\section{Diagonalization for asymmetric bands}\label{appendix_rotations}

Here we describe the full diagonalization procedure which holds for non-symmetric bands. Below the spin and charge indices have been suppressed, but the following applies to both the spin and charge sector (separately). If we have
\begin{eqnarray}
\begin{pmatrix}\phi_1(x)\\ \phi_2(x)\end{pmatrix}&=&
\begin{pmatrix} T^{\phi}_{11} & T^{\phi}_{12} \\ T^{\phi}_{21} & T^{\phi}_{22} \end{pmatrix}
\begin{pmatrix}\tphi_{1}(x)\\ \tphi_{2}(x)\end{pmatrix}\nonumber
\textrm{ and}\\
\begin{pmatrix}\theta_1(x)\\ \theta_2(x)\end{pmatrix}&=&
\begin{pmatrix} T^{\theta}_{11} & T^{\theta}_{12} \\ T^{\theta}_{21} & T^{\theta}_{22} \end{pmatrix}
\begin{pmatrix}\ttheta_{1}(x)\\ \ttheta_{2}(x)\end{pmatrix}.\label{rotate2}
\end{eqnarray}
then we require $[\mathbf{T}^{\phi}]^T=[\mathbf{T}^{\theta}]^{-1}$ for the canonical commutation relations to remain fulfilled for the transformed fields. There are several way to do this, the important point to note is that it can not be done with an orthogonal transformation which merely rotates the matrices.

We define $\mathbf{T}^{\phi}=\mathbf{P}\mathbf{\Lambda} \mathbf{Q}\mathbf{\tilde{\Lambda}}$ and $\mathbf{T}^{\theta}=\mathbf{P}\mathbf{\Lambda}^{-1} \mathbf{Q}\mathbf{\tilde{\Lambda}}^{-1}$, which automatically ensures that the commutation relations are held. $\mathbf{P}$ and $\mathbf{Q}$ are usual rotations (\emph{i.e.}~orthogonal matrices), and $\mathbf{\Lambda}$ and $\mathbf{\tilde{\Lambda}}$ are diagonal matrices which rescale the fields. The idea is that first we rotate and rescale such that  $\mathbf{\Lambda}\mathbf{P}^T\mathbf{M}'_\phi\mathbf{P}\mathbf{\Lambda}=\mathcal{I}$, the identity matrix. We then define $\mathbf{N}_\theta=\mathbf{\Lambda}\mathbf{P}^T\mathbf{M}'_\theta\mathbf{P}\mathbf{\Lambda}$, which is now a known symmetric matrix. This we diagonalize with $\mathbf{Q}$, which will of course leave the identity matrix unaffected. Finally we have a rescaling $\mathbf{\tilde{\Lambda}}$ so that $\ttheta$ and $\tphi$ have the same eigenvalues.

The two rescalings are $\mathbf{\Lambda}=\diag(\lambda^{-\frac{1}{2}}_1,\lambda^{-\frac{1}{2}}_2)$ and $\mathbf{\tilde{\Lambda}}=\diag(\tilde{\lambda}^{-\frac{1}{2}}_1,\tilde{\lambda}^{-\frac{1}{2}}_2)$ with
\begin{equation}
\lambda_{1,2}=\frac{v_1}{2K_1}+\frac{v_2}{2K_2}\pm\sqrt{v_B^2+\left(\frac{v_1}{2K_1}-\frac{v_2}{2K_2}\right)^2}
\end{equation}
and
\begin{equation}
\frac{1}{\tilde{\lambda}_{1,2}}=\sqrt{\frac{N^{\theta}_{11}}{2}+\frac{N^{\theta}_{22}}{2}\pm\sqrt{\left[N^{\theta}_{12}\right]^2+\left(\frac{N^{\theta}_{11}}{2}-\frac{N^{\theta}_{22}}{2}\right)^2}}\,.
\end{equation}
Therefore in the end $\mathbf{M}_\phi=\mathbf{M}_\theta=\diag[u_1,u_2]$ with
\begin{equation}
(u_{1,2})^2=\frac{N^{\theta}_{11}}{2}+\frac{N^{\theta}_{22}}{2}\pm\sqrt{\left[N^{\theta}_{12}\right]^2+\left(\frac{N^{\theta}_{11}}{2}-\frac{N^{\theta}_{22}}{2}\right)^2}\,.
\end{equation}
The Hamiltonian becomes
\begin{equation}
 H_q = \sum_{\substack{\nu = c,s\\\beta=1,2}}\frac{u_{\nu \beta}}{2} \int \ud x \left[ ( \partial_x \tphi^{\nu \beta} (x))^2 + ( \tPi^{\nu \beta} (x))^2 \right]\,.
\end{equation}
Note that contrary to the basis used in the main text, we have here rescaled the Luttinger parameters out of the Hamiltonian, equivalent to $\phi^{\nu\beta}\to\tphi^{\nu\beta}\sqrt{K^{\nu\beta}}$ and $\theta^{\nu\beta}\to\ttheta^{\nu\beta}/\sqrt{K^{\nu\beta}}$.

The generalization of the $SU(2)$ symmetry condition on the Luttinger parameters, see Sec.~\ref{sec_symmetry}, for asymmetric bands leads to the condition $\mathbf{T}^\phi_{s}=\mathbf{T}^\theta_{s}$, or equivalently $\mathbf{\Lambda}_{s} \mathbf{Q}_{s}\mathbf{\tilde{\Lambda}}_{s}=\mathbf{\Lambda}_{s}^{-1} \mathbf{Q}_{s}\mathbf{\tilde{\Lambda}}_{s}^{-1}$. This in turn tells us that $\lambda_{s1}=\lambda_{s2}=\tilde{\lambda}^{-1}_{s1}=\tilde{\lambda}^{-1}_{s2}$ and from this the conditions $v_{sA}=v_{sB}=0$, $K_{s1}=K_{s2}=1$, and $v_{s1}=v_{s2}$ follow directly. I.e.~for an $SU(2)$ symmetric system the spin part of the Hamiltonian is already diagonal in the band indices. It is therefore of course also diagonal in the $\beta=\pm$ basis.

\section{Expressions for velocities and Luttinger parameters}\label{app_param}

In this appendix we give low order expansions for the velocities in
terms of the bare interaction parameters and Fermi velocities. Note
that as before all kinematically indistinct processes are assumed to
be already appropriately rescaled.  The velocities and inter band
coupling terms are, for the spin sector,
\begin{eqnarray}
\label{A9}
v_{sA}&=&\frac{2\bar{g}_{4\parallel}-\bar{g}_{4\perp}-2\bar{g}_{2\parallel}+\bar{g}_{2\perp}}{16\pi}\,,\nonumber\\
v_{sB}&=&\frac{2\bar{g}_{4\parallel}+\bar{g}_{4\perp}-2\bar{g}_{2\parallel}-\bar{g}_{2\perp}}{16\pi}\,,\textrm{ and}\\
v^2_{sb}&=&\left(v_{Fb}-\frac{2g_{2b\parallel}-g_{2b\perp}-4g_{4\parallel b}+2g_{4\perp b}}{8\pi}\right)\nonumber\\ &&\times
\left(v_{Fb}-\frac{2g_{2b\parallel}+g_{2b\perp}-4g_{4\parallel b}-2g_{4\perp b}}{8\pi}\right)\,\nonumber
\end{eqnarray}
with band index $b=1,2$. In the charge sector we have
\begin{eqnarray}
\label{A10}
v_{cA}&=&\frac{2\bar{g}_{4\parallel}-\bar{g}_{4\perp}+2\bar{g}_{2\parallel}-\bar{g}_{2\perp}}{16\pi},\nonumber\\
v_{cB}&=&\frac{2\bar{g}_{4\parallel}+\bar{g}_{4\perp}+2\bar{g}_{2\parallel}+\bar{g}_{2\perp}}{16\pi}\textrm{, and}\\
v^2_{cb}&=&\left(v_{Fb}+\frac{2g_{2b\parallel}-g_{2b\perp}+4g_{4\parallel b}-2g_{4\perp b}}{8\pi}\right)\nonumber\\ &&\times
\left(v_{Fb}+\frac{2g_{2b\parallel}+g_{2b\perp}+4g_{4\parallel b}+2g_{4\perp b}}{8\pi}\right)\,.\nonumber
\end{eqnarray}
The Luttinger parameters are
\begin{eqnarray}
K_{sb}^2&=&\left(v_{Fb}-\frac{2g_{2b\parallel}-g_{2b\perp}-4g_{4\parallel b}+2g_{4\perp b}}{8\pi}\right)\\ &&\times
\left(v_{Fb}-\frac{2g_{2b\parallel}+g_{2b\perp}-4g_{4\parallel b}-2g_{4\perp b}}{8\pi}\right)^{-1}\nonumber
\end{eqnarray}
and for the charge sector
\begin{eqnarray}
K_{cb}^2&=&\left(v_{Fb}+\frac{2g_{2b\parallel}-g_{2b\perp}+4g_{4\parallel b}-2g_{4\perp b}}{8\pi}\right)\\ &&\times
\left(v_{Fb}+\frac{2g_{2b\parallel}+g_{2b\perp}+4g_{4\parallel b}+2g_{4\perp b}}{8\pi}\right)^{-1}\,.\nonumber
\end{eqnarray}
The multitude of $g$ parameters will depend on the specific microscopic model under consideration.

For the symmetric band model which we consider in the main text we find, with $v_{\nu b}=v_{\nu}$ and $K_{\nu b}=K_{\nu}$,
\begin{eqnarray}
(u^{\nu\pm})^2&=&\left(v_\nu\pm v_{\nu B}K_\nu\right)\left(v_\nu\pm v_{\nu A}/K_\nu\right)
\textrm{, and}\nonumber\\
\frac{1}{(K^{\nu\pm})^2}&=&\frac{1}{K_\nu^2}\frac{v_\nu\pm v_{\nu B}K_\nu}{v_\nu\pm v_{\nu A}/K_\nu}\,,
\end{eqnarray}
for the velocities and Luttinger parameters of the normal modes.

\section{Interactions with finite momentum transfer}\label{appendix_umklapp}

At special fillings where the total transferred momentum in a scattering process becomes commensurate with the lattice, additional interactions to those considered in section \ref{sec_model} can become important. For completeness we list them here. For $2(k_{F1}+k_{F2})=2\pi$ we have the following additional processes which we have neglected. Firstly
\begin{eqnarray}
\bar{H}_{1}&=&\sum_{\sigma,\sigma',b,r}\int\ud x\frac{\bar{g}_{1}}{2}
\psi_{\sigma' \bar{b}r}^\dagger\psi_{\sigma' \bar{b}\bar{r}}\psi_{\sigma b\bar{r}}^\dagger
\psi_{\sigma br}\,,\nonumber\\
\bar{H}'_{2\perp}&=&\sum_{\sigma,b,r}\int\ud x\frac{\bar{g}'_{2\perp}}{2}
\psi_{\bar{\sigma} b\bar{r}}^\dagger\psi_{\bar{\sigma} \bar{b}\bar{r}}\psi_{\sigma \bar{b}r}^\dagger
\psi_{\sigma br}\,,\\
H'_{4\perp}&=&\sum_{\sigma,\sigma',b,r}\int\ud x\frac{g'_{4\perp}}{2}
\psi_{\sigma' \bar{b}r}^\dagger\psi_{\sigma' br}\psi_{\sigma \bar{b}r}^\dagger
\psi_{\sigma br}\,.\nonumber
\end{eqnarray}
Bosonized they are, firstly,
\begin{eqnarray}
\bar{H}_{1\parallel}&=&\frac{\bar{g}_{1\parallel}}{(\pi \alpha)^2}\int\ud x\cos\left[\sqrt{4\pi}\phi^{s-}\right]\cos\left[\sqrt{4\pi}\phi^{c-}\right]\,,\nonumber\\
\bar{H}_{1\perp}&=&\frac{\bar{g}_{1\perp}}{(\pi \alpha)^2}\int\ud x\cos\left[\sqrt{4\pi}\phi^{s+}\right]\cos\left[\sqrt{4\pi}\phi^{c-}\right]\,,\qquad\\
\bar{H}'_{2\perp}&=&\frac{\bar{g}'_{2\perp}}{(\pi \alpha)^2}\int\ud x\cos\left[\sqrt{4\pi}\theta^{s-}\right]\cos\left[\sqrt{4\pi}\phi^{c-}\right]\,.\qquad\nonumber
\end{eqnarray}
Finally
\begin{eqnarray}
H'_{4\parallel}&=&\frac{g'_{4\parallel}}{(\pi \alpha)^2}\int\ud x\times\\&&\nonumber\left\{\prod_\nu\cos \left[ \sqrt{4\pi} \phi^{\nu-} \right]\right.
\cos \left[ \sqrt{4\pi} \theta^{s-} \right]\cos \left[\sqrt{4\pi} \theta^{c-} \right]\\\nonumber\
&&+\prod_\nu\sin \left[ \sqrt{4\pi} \phi^{\nu-} \right]
\left.
\sin \left[ \sqrt{4\pi} \theta^{s-} \right]\sin \left[\sqrt{4\pi} \theta^{c-} \right]\right\}\\\nonumber
H'_{4\perp}&=&\frac{g'_{4\perp}}{(\pi \alpha)^2}\int\ud x\cos\left[\sqrt{4\pi}\phi^{c-}\right]\cos\left[\sqrt{4\pi}\theta^{c-}\right]\,.\qquad
\end{eqnarray}

There is also a particular umklapp process for which the oscillations can become commensurate with the lattice. This is a $3k_{F2}-k_{F1}$ momentum transfer processes:
\begin{equation}
H_{U}=g_{U}\sum_{\sigma,\sigma',b,r}\int\!\!\ud x\e^{-ir(3k_{F2}-k_{F1})x}
\psi_{\sigma' \bar{b}\bar{r}}^\dagger\psi_{\sigma' br}\psi_{\sigma 2\bar{r}}^\dagger\psi_{\sigma 2r}\,.
\end{equation}
In the bosonic form this becomes
\begin{eqnarray}
H_{U}=\frac{g_{U}}{(\pi \alpha)^2}\sum_\sigma\int\ud x\cos\left[\sqrt{\pi}\left(\theta^{c -}-\sigma\theta^{s -}\right)\right]\times\nonumber\\
\cos\left[\left(3k_{F2}-k_{F1}\right)x-\sqrt{\pi}\left(\sqrt{4}\phi^{c+}-\phi^{c -}-\sigma\phi^{s -}\right)\right]\,.\nonumber\\{}
\end{eqnarray}
In particular, though not solely, one can see that this umklapp scattering term will become important for the highly symmetric scenario $k_{F2}=3k_{F1}=3\pi/4$ as then $3k_{F2}-k_{F1}=2\pi$. In this case all of these momentum transfer processes listed in this appendix will be present.

The scaling dimensions of these interactions necessary for the first order RG equations are, in the band symmetric model:
\begin{eqnarray}
\bar{\gamma}_{1\parallel}&=&
K^{s-}+K^{c-}\,,\quad
\bar{\gamma}_{1\perp}=
K^{s+}+K^{c+}\,,\nonumber\\
\bar{\gamma}'_{2\perp}&=&
K^{c-}+\frac{1}{K^{s-}}\,,\quad
\gamma'_{4\perp}=
K^{c-}+\frac{1}{K^{c-}}\,,\nonumber\\
\gamma'_{4\parallel}&=&
K^{s-}+K^{c-}+\frac{1}{K^{s-}}+\frac{1}{K^{c-}}\,,\\
\gamma_U&=&K^{c+}+\frac{1}{4}\left[K^{s-}+K^{c-}+\frac{1}{K^{s-}}+\frac{1}{K^{c-}}\right]\,.\nonumber
\end{eqnarray}
$g_{4\parallel}'$ is irrelevant and $g_{4\perp}'$ is at best marginal. In the general non $SU(2)$ symmetric case we consider the rest would have to be taken into account. In principle all of these processes can also show up in the double and single spin-flip versions as well. An exhaustive list of all these possibilities and their influence on the phase diagram is beyond the scope of this paper.


\begin{thebibliography}{52}
\expandafter\ifx\csname natexlab\endcsname\relax\def\natexlab#1{#1}\fi
\expandafter\ifx\csname bibnamefont\endcsname\relax
  \def\bibnamefont#1{#1}\fi
\expandafter\ifx\csname bibfnamefont\endcsname\relax
  \def\bibfnamefont#1{#1}\fi
\expandafter\ifx\csname citenamefont\endcsname\relax
  \def\citenamefont#1{#1}\fi
\expandafter\ifx\csname url\endcsname\relax
  \def\url#1{\texttt{#1}}\fi
\expandafter\ifx\csname urlprefix\endcsname\relax\def\urlprefix{URL }\fi
\providecommand{\bibinfo}[2]{#2}
\providecommand{\eprint}[2][]{\url{#2}}

\bibitem[{\citenamefont{Haldane}(1981)}]{Haldane1981a}
\bibinfo{author}{\bibfnamefont{F.~D.~M.} \bibnamefont{Haldane}},
  \bibinfo{journal}{Journal of Physics C: Solid State Physics}
  \textbf{\bibinfo{volume}{14}}, \bibinfo{pages}{2585} (\bibinfo{year}{1981}).

\bibitem[{\citenamefont{Giamarchi}(2004)}]{Giamarchi2004}
\bibinfo{author}{\bibfnamefont{T.}~\bibnamefont{Giamarchi}},
  \emph{\bibinfo{title}{Quantum Physics in One Dimension}}
  (\bibinfo{publisher}{Clarendon Press, Oxford}, \bibinfo{year}{2004}).

\bibitem[{\citenamefont{Motoyama et~al.}(1996)\citenamefont{Motoyama, Eisaki,
  and Uchida}}]{MotoyamaEisaki}
\bibinfo{author}{\bibfnamefont{N.}~\bibnamefont{Motoyama}},
  \bibinfo{author}{\bibfnamefont{H.}~\bibnamefont{Eisaki}}, \bibnamefont{and}
  \bibinfo{author}{\bibfnamefont{S.}~\bibnamefont{Uchida}},
  \bibinfo{journal}{Phys. Rev. Lett.} \textbf{\bibinfo{volume}{76}},
  \bibinfo{pages}{3212} (\bibinfo{year}{1996}).

\bibitem[{\citenamefont{Hase et~al.}(1993)\citenamefont{Hase, Terasaki, and
  Uchinokura}}]{Hase1993a}
\bibinfo{author}{\bibfnamefont{M.}~\bibnamefont{Hase}},
  \bibinfo{author}{\bibfnamefont{I.}~\bibnamefont{Terasaki}}, \bibnamefont{and}
  \bibinfo{author}{\bibfnamefont{K.}~\bibnamefont{Uchinokura}},
  \bibinfo{journal}{Phys. Rev. Lett.} \textbf{\bibinfo{volume}{70}},
  \bibinfo{pages}{3651} (\bibinfo{year}{1993}).

\bibitem[{\citenamefont{Dender et~al.}(1997)\citenamefont{Dender, Hammar,
  Reich, Broholm, and Aeppli}}]{Dender1997}
\bibinfo{author}{\bibfnamefont{D.~C.} \bibnamefont{Dender}},
  \bibinfo{author}{\bibfnamefont{P.~R.} \bibnamefont{Hammar}},
  \bibinfo{author}{\bibfnamefont{D.~H.} \bibnamefont{Reich}},
  \bibinfo{author}{\bibfnamefont{C.}~\bibnamefont{Broholm}}, \bibnamefont{and}
  \bibinfo{author}{\bibfnamefont{G.}~\bibnamefont{Aeppli}},
  \bibinfo{journal}{Phys. Rev. Lett.} \textbf{\bibinfo{volume}{79}},
  \bibinfo{pages}{1750} (\bibinfo{year}{1997}).

\bibitem[{\citenamefont{Eggert and Affleck}(1992)}]{EggertAffleck92}
\bibinfo{author}{\bibfnamefont{S.}~\bibnamefont{Eggert}} \bibnamefont{and}
  \bibinfo{author}{\bibfnamefont{I.}~\bibnamefont{Affleck}},
  \bibinfo{journal}{Phys. Rev. B} \textbf{\bibinfo{volume}{46}},
  \bibinfo{pages}{10866} (\bibinfo{year}{1992}).

\bibitem[{\citenamefont{Eggert et~al.}(1994)\citenamefont{Eggert, Affleck, and
  Takahashi}}]{egg94}
\bibinfo{author}{\bibfnamefont{S.}~\bibnamefont{Eggert}},
  \bibinfo{author}{\bibfnamefont{I.}~\bibnamefont{Affleck}}, \bibnamefont{and}
  \bibinfo{author}{\bibfnamefont{M.}~\bibnamefont{Takahashi}},
  \bibinfo{journal}{Phys. Rev. Lett.} \textbf{\bibinfo{volume}{73}},
  \bibinfo{pages}{332} (\bibinfo{year}{1994}).

\bibitem[{\citenamefont{Oshikawa and Affleck}(1997)}]{OshikawaAffleck}
\bibinfo{author}{\bibfnamefont{M.}~\bibnamefont{Oshikawa}} \bibnamefont{and}
  \bibinfo{author}{\bibfnamefont{I.}~\bibnamefont{Affleck}},
  \bibinfo{journal}{Phys. Rev. Lett.} \textbf{\bibinfo{volume}{79}},
  \bibinfo{pages}{2883} (\bibinfo{year}{1997}).

\bibitem[{\citenamefont{Sirker et~al.}(2007)\citenamefont{Sirker, Laflorencie,
  Fujimoto, Eggert, and Affleck}}]{SirkerLaflorencie}
\bibinfo{author}{\bibfnamefont{J.}~\bibnamefont{Sirker}},
  \bibinfo{author}{\bibfnamefont{N.}~\bibnamefont{Laflorencie}},
  \bibinfo{author}{\bibfnamefont{S.}~\bibnamefont{Fujimoto}},
  \bibinfo{author}{\bibfnamefont{S.}~\bibnamefont{Eggert}}, \bibnamefont{and}
  \bibinfo{author}{\bibfnamefont{I.}~\bibnamefont{Affleck}},
  \bibinfo{journal}{Phys. Rev. Lett.} \textbf{\bibinfo{volume}{98}},
  \bibinfo{pages}{137205} (\bibinfo{year}{2007}).

\bibitem[{\citenamefont{Sirker et~al.}(2008)\citenamefont{Sirker, Laflorencie,
  Fujimoto, Eggert, and Affleck}}]{SirkerLaflorencie2}
\bibinfo{author}{\bibfnamefont{J.}~\bibnamefont{Sirker}},
  \bibinfo{author}{\bibfnamefont{N.}~\bibnamefont{Laflorencie}},
  \bibinfo{author}{\bibfnamefont{S.}~\bibnamefont{Fujimoto}},
  \bibinfo{author}{\bibfnamefont{S.}~\bibnamefont{Eggert}}, \bibnamefont{and}
  \bibinfo{author}{\bibfnamefont{I.}~\bibnamefont{Affleck}},
  \bibinfo{journal}{J. Stat. Mech.} \bibinfo{pages}{P02015}
  (\bibinfo{year}{2008}).

\bibitem[{\citenamefont{Sirker et~al.}(2009)\citenamefont{Sirker, Pereira, and
  Affleck}}]{SirkerPereira}
\bibinfo{author}{\bibfnamefont{J.}~\bibnamefont{Sirker}},
  \bibinfo{author}{\bibfnamefont{R.~G.} \bibnamefont{Pereira}},
  \bibnamefont{and} \bibinfo{author}{\bibfnamefont{I.}~\bibnamefont{Affleck}},
  \bibinfo{journal}{Phys. Rev. Lett.} \textbf{\bibinfo{volume}{103}},
  \bibinfo{pages}{216602} (\bibinfo{year}{2009}).

\bibitem[{\citenamefont{Sirker et~al.}(2011)\citenamefont{Sirker, Pereira, and
  Affleck}}]{SirkerPereira2}
\bibinfo{author}{\bibfnamefont{J.}~\bibnamefont{Sirker}},
  \bibinfo{author}{\bibfnamefont{R.~G.} \bibnamefont{Pereira}},
  \bibnamefont{and} \bibinfo{author}{\bibfnamefont{I.}~\bibnamefont{Affleck}},
  \bibinfo{journal}{Phys. Rev. B} \textbf{\bibinfo{volume}{83}},
  \bibinfo{pages}{035115} (\bibinfo{year}{2011}).

\bibitem[{\citenamefont{Pereira et~al.}(2006)\citenamefont{Pereira, Sirker,
  Caux, Hagemans, Maillet, White, and Affleck}}]{PereiraSirker}
\bibinfo{author}{\bibfnamefont{R.~G.} \bibnamefont{Pereira}},
  \bibinfo{author}{\bibfnamefont{J.}~\bibnamefont{Sirker}},
  \bibinfo{author}{\bibfnamefont{J.-S.} \bibnamefont{Caux}},
  \bibinfo{author}{\bibfnamefont{R.}~\bibnamefont{Hagemans}},
  \bibinfo{author}{\bibfnamefont{J.~M.} \bibnamefont{Maillet}},
  \bibinfo{author}{\bibfnamefont{S.~R.} \bibnamefont{White}}, \bibnamefont{and}
  \bibinfo{author}{\bibfnamefont{I.}~\bibnamefont{Affleck}},
  \bibinfo{journal}{Phys. Rev. Lett.} \textbf{\bibinfo{volume}{96}},
  \bibinfo{pages}{257202} (\bibinfo{year}{2006}).

\bibitem[{\citenamefont{Pereira et~al.}(2007)\citenamefont{Pereira, Sirker,
  Caux, Hagemans, Maillet, White, and Affleck}}]{PereiraSirkerJSTAT}
\bibinfo{author}{\bibfnamefont{R.~G.} \bibnamefont{Pereira}},
  \bibinfo{author}{\bibfnamefont{J.}~\bibnamefont{Sirker}},
  \bibinfo{author}{\bibfnamefont{J.-S.} \bibnamefont{Caux}},
  \bibinfo{author}{\bibfnamefont{R.}~\bibnamefont{Hagemans}},
  \bibinfo{author}{\bibfnamefont{J.~M.} \bibnamefont{Maillet}},
  \bibinfo{author}{\bibfnamefont{S.~R.} \bibnamefont{White}}, \bibnamefont{and}
  \bibinfo{author}{\bibfnamefont{I.}~\bibnamefont{Affleck}},
  \bibinfo{journal}{J. Stat. Mech.} \bibinfo{pages}{P08022}
  (\bibinfo{year}{2007}).

\bibitem[{\citenamefont{Auslaender et~al.}(2005)\citenamefont{Auslaender,
  Steinberg, Yacoby, Tserkovnyak, Halperin, Baldwin, Pfeiffer, and
  West}}]{AuslaenderSteinberg}
\bibinfo{author}{\bibfnamefont{O.~M.} \bibnamefont{Auslaender}},
  \bibinfo{author}{\bibfnamefont{H.}~\bibnamefont{Steinberg}},
  \bibinfo{author}{\bibfnamefont{A.}~\bibnamefont{Yacoby}},
  \bibinfo{author}{\bibfnamefont{Y.}~\bibnamefont{Tserkovnyak}},
  \bibinfo{author}{\bibfnamefont{B.~I.} \bibnamefont{Halperin}},
  \bibinfo{author}{\bibfnamefont{K.~W.} \bibnamefont{Baldwin}},
  \bibinfo{author}{\bibfnamefont{L.~N.} \bibnamefont{Pfeiffer}},
  \bibnamefont{and} \bibinfo{author}{\bibfnamefont{K.~W.} \bibnamefont{West}},
  \bibinfo{journal}{Science} \textbf{\bibinfo{volume}{308}},
  \bibinfo{pages}{88} (\bibinfo{year}{2005}).

\bibitem[{\citenamefont{Yao et~al.}(1999)\citenamefont{Yao, Postma, Balents,
  and Dekker}}]{YaoPostma}
\bibinfo{author}{\bibfnamefont{Z.}~\bibnamefont{Yao}},
  \bibinfo{author}{\bibfnamefont{H.~W.~C.} \bibnamefont{Postma}},
  \bibinfo{author}{\bibfnamefont{L.}~\bibnamefont{Balents}}, \bibnamefont{and}
  \bibinfo{author}{\bibfnamefont{C.}~\bibnamefont{Dekker}},
  \bibinfo{journal}{Nature} \textbf{\bibinfo{volume}{402}},
  \bibinfo{pages}{273} (\bibinfo{year}{1999}).

\bibitem[{\citenamefont{Bockrath et~al.}(1999)\citenamefont{Bockrath, Cobden,
  Lu, Rinzler, Smalley, Balents, and McEuen}}]{Bockrath}
\bibinfo{author}{\bibfnamefont{M.}~\bibnamefont{Bockrath}},
  \bibinfo{author}{\bibfnamefont{D.~H.} \bibnamefont{Cobden}},
  \bibinfo{author}{\bibfnamefont{J.}~\bibnamefont{Lu}},
  \bibinfo{author}{\bibfnamefont{A.~G.} \bibnamefont{Rinzler}},
  \bibinfo{author}{\bibfnamefont{R.~E.} \bibnamefont{Smalley}},
  \bibinfo{author}{\bibfnamefont{L.}~\bibnamefont{Balents}}, \bibnamefont{and}
  \bibinfo{author}{\bibfnamefont{P.~L.} \bibnamefont{McEuen}},
  \bibinfo{journal}{Nature} \textbf{\bibinfo{volume}{397}},
  \bibinfo{pages}{598} (\bibinfo{year}{1999}).

\bibitem[{\citenamefont{Jompol et~al.}(2009)\citenamefont{Jompol, Ford,
  Griffiths, Farrer, Jones, Anderson, Ritchie, Silk, and
  Schofield}}]{Jompol2009}
\bibinfo{author}{\bibfnamefont{Y.}~\bibnamefont{Jompol}},
  \bibinfo{author}{\bibfnamefont{C.~J.~B.} \bibnamefont{Ford}},
  \bibinfo{author}{\bibfnamefont{J.~P.} \bibnamefont{Griffiths}},
  \bibinfo{author}{\bibfnamefont{I.}~\bibnamefont{Farrer}},
  \bibinfo{author}{\bibfnamefont{G.~A.~C.} \bibnamefont{Jones}},
  \bibinfo{author}{\bibfnamefont{D.}~\bibnamefont{Anderson}},
  \bibinfo{author}{\bibfnamefont{D.~A.} \bibnamefont{Ritchie}},
  \bibinfo{author}{\bibfnamefont{T.~W.} \bibnamefont{Silk}}, \bibnamefont{and}
  \bibinfo{author}{\bibfnamefont{A.~J.} \bibnamefont{Schofield}},
  \bibinfo{journal}{Science} \textbf{\bibinfo{volume}{325}},
  \bibinfo{pages}{597} (\bibinfo{year}{2009}).

\bibitem[{\citenamefont{Deshpande et~al.}(2010)\citenamefont{Deshpande,
  Bockrath, Glazman, and Yacoby}}]{Deshpande2010}
\bibinfo{author}{\bibfnamefont{V.~V.} \bibnamefont{Deshpande}},
  \bibinfo{author}{\bibfnamefont{M.}~\bibnamefont{Bockrath}},
  \bibinfo{author}{\bibfnamefont{L.~I.} \bibnamefont{Glazman}},
  \bibnamefont{and} \bibinfo{author}{\bibfnamefont{A.}~\bibnamefont{Yacoby}},
  \bibinfo{journal}{Nature} \textbf{\bibinfo{volume}{464}},
  \bibinfo{pages}{209} (\bibinfo{year}{2010}).

\bibitem[{\citenamefont{Segovia et~al.}(1999)\citenamefont{Segovia, Purdie,
  Hengsberger, and Baer}}]{SegoviaPurdie}
\bibinfo{author}{\bibfnamefont{P.}~\bibnamefont{Segovia}},
  \bibinfo{author}{\bibfnamefont{D.}~\bibnamefont{Purdie}},
  \bibinfo{author}{\bibfnamefont{M.}~\bibnamefont{Hengsberger}},
  \bibnamefont{and} \bibinfo{author}{\bibfnamefont{Y.}~\bibnamefont{Baer}},
  \bibinfo{journal}{Nature} \textbf{\bibinfo{volume}{402}},
  \bibinfo{pages}{504} (\bibinfo{year}{1999}).

\bibitem[{\citenamefont{Losio et~al.}(2001)\citenamefont{Losio, Altmann,
  Kirakosian, Lin, Petrovykh, and Himpsel}}]{Losio2001}
\bibinfo{author}{\bibfnamefont{R.}~\bibnamefont{Losio}},
  \bibinfo{author}{\bibfnamefont{K.~N.} \bibnamefont{Altmann}},
  \bibinfo{author}{\bibfnamefont{A.}~\bibnamefont{Kirakosian}},
  \bibinfo{author}{\bibfnamefont{J.-L.} \bibnamefont{Lin}},
  \bibinfo{author}{\bibfnamefont{D.~Y.} \bibnamefont{Petrovykh}},
  \bibnamefont{and} \bibinfo{author}{\bibfnamefont{F.~J.}
  \bibnamefont{Himpsel}}, \bibinfo{journal}{Phys. Rev. Lett.}
  \textbf{\bibinfo{volume}{86}}, \bibinfo{pages}{4632} (\bibinfo{year}{2001}).

\bibitem[{\citenamefont{Ahn et~al.}(2003)\citenamefont{Ahn, Yeom, Yoon, and
  Lyo}}]{Ahn2003}
\bibinfo{author}{\bibfnamefont{J.~R.} \bibnamefont{Ahn}},
  \bibinfo{author}{\bibfnamefont{H.~W.} \bibnamefont{Yeom}},
  \bibinfo{author}{\bibfnamefont{H.~S.} \bibnamefont{Yoon}}, \bibnamefont{and}
  \bibinfo{author}{\bibfnamefont{I.-W.} \bibnamefont{Lyo}},
  \bibinfo{journal}{Phys. Rev. Lett.} \textbf{\bibinfo{volume}{91}},
  \bibinfo{pages}{196403} (\bibinfo{year}{2003}).

\bibitem[{\citenamefont{S\'anchez-Portal
  et~al.}(2004)\citenamefont{S\'anchez-Portal, Riikonen, and
  Martin}}]{S'anchez-Portal2004}
\bibinfo{author}{\bibfnamefont{D.}~\bibnamefont{S\'anchez-Portal}},
  \bibinfo{author}{\bibfnamefont{S.}~\bibnamefont{Riikonen}}, \bibnamefont{and}
  \bibinfo{author}{\bibfnamefont{R.~M.} \bibnamefont{Martin}},
  \bibinfo{journal}{Phys. Rev. Lett.} \textbf{\bibinfo{volume}{93}},
  \bibinfo{pages}{146803} (\bibinfo{year}{2004}).

\bibitem[{\citenamefont{Barke et~al.}(2006)\citenamefont{Barke, Zheng,
  R\"ugheimer, and Himpsel}}]{Barke2006}
\bibinfo{author}{\bibfnamefont{I.}~\bibnamefont{Barke}},
  \bibinfo{author}{\bibfnamefont{F.}~\bibnamefont{Zheng}},
  \bibinfo{author}{\bibfnamefont{T.~K.} \bibnamefont{R\"ugheimer}},
  \bibnamefont{and} \bibinfo{author}{\bibfnamefont{F.~J.}
  \bibnamefont{Himpsel}}, \bibinfo{journal}{Phys. Rev. Lett.}
  \textbf{\bibinfo{volume}{97}}, \bibinfo{pages}{226405}
  (\bibinfo{year}{2006}).

\bibitem[{\citenamefont{Blumenstein et~al.}(2011)\citenamefont{Blumenstein,
  Sch\"affer, Mietke, Meyer, Dollinger, Lochner, Cui, Patthey, Matzdorf, and
  Claessen}}]{Blumenstein2011}
\bibinfo{author}{\bibfnamefont{C.}~\bibnamefont{Blumenstein}},
  \bibinfo{author}{\bibfnamefont{J.}~\bibnamefont{Sch\"affer}},
  \bibinfo{author}{\bibfnamefont{S.}~\bibnamefont{Mietke}},
  \bibinfo{author}{\bibfnamefont{S.}~\bibnamefont{Meyer}},
  \bibinfo{author}{\bibfnamefont{A.}~\bibnamefont{Dollinger}},
  \bibinfo{author}{\bibfnamefont{M.}~\bibnamefont{Lochner}},
  \bibinfo{author}{\bibfnamefont{X.}~\bibnamefont{Cui}},
  \bibinfo{author}{\bibfnamefont{L.}~\bibnamefont{Patthey}},
  \bibinfo{author}{\bibfnamefont{R.}~\bibnamefont{Matzdorf}}, \bibnamefont{and}
  \bibinfo{author}{\bibfnamefont{R.}~\bibnamefont{Claessen}},
  \bibinfo{journal}{Nat. Phys.} \textbf{\bibinfo{volume}{7}},
  \bibinfo{pages}{776} (\bibinfo{year}{2011}).

\bibitem[{\citenamefont{Sch\"afer et~al.}(2008)\citenamefont{Sch\"afer,
  Blumenstein, Meyer, Wisniewski, and Claessen}}]{Schafer2008}
\bibinfo{author}{\bibfnamefont{J.}~\bibnamefont{Sch\"afer}},
  \bibinfo{author}{\bibfnamefont{C.}~\bibnamefont{Blumenstein}},
  \bibinfo{author}{\bibfnamefont{S.}~\bibnamefont{Meyer}},
  \bibinfo{author}{\bibfnamefont{M.}~\bibnamefont{Wisniewski}},
  \bibnamefont{and} \bibinfo{author}{\bibfnamefont{R.}~\bibnamefont{Claessen}},
  \bibinfo{journal}{Phys. Rev. Lett.} \textbf{\bibinfo{volume}{101}},
  \bibinfo{pages}{236802} (\bibinfo{year}{2008}).

\bibitem[{\citenamefont{Meyer et~al.}(2011)\citenamefont{Meyer, Sch\"afer,
  Blumenstein, H\"opfner, Bostwick, McChesney, Rotenberg, and
  Claessen}}]{Meyer2011}
\bibinfo{author}{\bibfnamefont{S.}~\bibnamefont{Meyer}},
  \bibinfo{author}{\bibfnamefont{J.}~\bibnamefont{Sch\"afer}},
  \bibinfo{author}{\bibfnamefont{C.}~\bibnamefont{Blumenstein}},
  \bibinfo{author}{\bibfnamefont{P.}~\bibnamefont{H\"opfner}},
  \bibinfo{author}{\bibfnamefont{A.}~\bibnamefont{Bostwick}},
  \bibinfo{author}{\bibfnamefont{J.~L.} \bibnamefont{McChesney}},
  \bibinfo{author}{\bibfnamefont{E.}~\bibnamefont{Rotenberg}},
  \bibnamefont{and} \bibinfo{author}{\bibfnamefont{R.}~\bibnamefont{Claessen}},
  \bibinfo{journal}{Phys. Rev. B} \textbf{\bibinfo{volume}{83}},
  \bibinfo{pages}{121411} (\bibinfo{year}{2011}).

\bibitem[{\citenamefont{Winkler}(2003)}]{Winkler2003}
\bibinfo{author}{\bibfnamefont{R.}~\bibnamefont{Winkler}},
  \emph{\bibinfo{title}{Spin-Orbit Coupling Effects in Two-Dimensional Electron
  and Hole Systems}} (\bibinfo{publisher}{Springer, Berlin},
  \bibinfo{year}{2003}).

\bibitem[{\citenamefont{H\"opfner et~al.}(2012)\citenamefont{H\"opfner,
  Sch\"afer, Fleszar, Dil, Slomski, Meier, Loho, Blumenstein, Patthey, Hanke
  et~al.}}]{HoepfnerSchaefer}
\bibinfo{author}{\bibfnamefont{P.}~\bibnamefont{H\"opfner}},
  \bibinfo{author}{\bibfnamefont{J.}~\bibnamefont{Sch\"afer}},
  \bibinfo{author}{\bibfnamefont{A.}~\bibnamefont{Fleszar}},
  \bibinfo{author}{\bibfnamefont{J.~H.} \bibnamefont{Dil}},
  \bibinfo{author}{\bibfnamefont{B.}~\bibnamefont{Slomski}},
  \bibinfo{author}{\bibfnamefont{F.}~\bibnamefont{Meier}},
  \bibinfo{author}{\bibfnamefont{C.}~\bibnamefont{Loho}},
  \bibinfo{author}{\bibfnamefont{C.}~\bibnamefont{Blumenstein}},
  \bibinfo{author}{\bibfnamefont{L.}~\bibnamefont{Patthey}},
  \bibinfo{author}{\bibfnamefont{W.}~\bibnamefont{Hanke}},
  \bibnamefont{et~al.}, \bibinfo{journal}{Phys. Rev. Lett.}
  \textbf{\bibinfo{volume}{108}}, \bibinfo{pages}{186801}
  (\bibinfo{year}{2012}).

\bibitem[{\citenamefont{Schulz et~al.}(2010)\citenamefont{Schulz, De~Martino,
  and Egger}}]{Schulz2010}
\bibinfo{author}{\bibfnamefont{A.}~\bibnamefont{Schulz}},
  \bibinfo{author}{\bibfnamefont{A.}~\bibnamefont{De~Martino}},
  \bibnamefont{and} \bibinfo{author}{\bibfnamefont{R.}~\bibnamefont{Egger}},
  \bibinfo{journal}{Phys. Rev. B} \textbf{\bibinfo{volume}{82}},
  \bibinfo{pages}{033407} (\bibinfo{year}{2010}).

\bibitem[{\citenamefont{Gangadharaiah et~al.}(2008)\citenamefont{Gangadharaiah,
  Sun, and Starykh}}]{Gangadharaiah2008}
\bibinfo{author}{\bibfnamefont{S.}~\bibnamefont{Gangadharaiah}},
  \bibinfo{author}{\bibfnamefont{J.}~\bibnamefont{Sun}}, \bibnamefont{and}
  \bibinfo{author}{\bibfnamefont{O.~A.} \bibnamefont{Starykh}},
  \bibinfo{journal}{Phys. Rev. B} \textbf{\bibinfo{volume}{78}},
  \bibinfo{pages}{054436} (\bibinfo{year}{2008}).

\bibitem[{\citenamefont{Schulz et~al.}(2009)\citenamefont{Schulz, De~Martino,
  Ingenhoven, and Egger}}]{Schulz2009}
\bibinfo{author}{\bibfnamefont{A.}~\bibnamefont{Schulz}},
  \bibinfo{author}{\bibfnamefont{A.}~\bibnamefont{De~Martino}},
  \bibinfo{author}{\bibfnamefont{P.}~\bibnamefont{Ingenhoven}},
  \bibnamefont{and} \bibinfo{author}{\bibfnamefont{R.}~\bibnamefont{Egger}},
  \bibinfo{journal}{Phys. Rev. B} \textbf{\bibinfo{volume}{79}},
  \bibinfo{pages}{205432} (\bibinfo{year}{2009}).

\bibitem[{\citenamefont{Varma and Zawadowski}(1985)}]{Varma1985}
\bibinfo{author}{\bibfnamefont{C.~M.} \bibnamefont{Varma}} \bibnamefont{and}
  \bibinfo{author}{\bibfnamefont{A.}~\bibnamefont{Zawadowski}},
  \bibinfo{journal}{Phys. Rev. B} \textbf{\bibinfo{volume}{32}},
  \bibinfo{pages}{7399} (\bibinfo{year}{1985}).

\bibitem[{\citenamefont{Penc and S\'olyom}(1990)}]{Penc1990}
\bibinfo{author}{\bibfnamefont{K.}~\bibnamefont{Penc}} \bibnamefont{and}
  \bibinfo{author}{\bibfnamefont{J.}~\bibnamefont{S\'olyom}},
  \bibinfo{journal}{Phys. Rev. B} \textbf{\bibinfo{volume}{41}},
  \bibinfo{pages}{704} (\bibinfo{year}{1990}).

\bibitem[{\citenamefont{Finkel'stein and Larkin}(1993)}]{Finkel'stein1993}
\bibinfo{author}{\bibfnamefont{A.~M.} \bibnamefont{Finkel'stein}}
  \bibnamefont{and} \bibinfo{author}{\bibfnamefont{A.~I.}
  \bibnamefont{Larkin}}, \bibinfo{journal}{Phys. Rev. B}
  \textbf{\bibinfo{volume}{47}}, \bibinfo{pages}{10461} (\bibinfo{year}{1993}).

\bibitem[{\citenamefont{Fabrizio}(1993)}]{Fabrizio1993}
\bibinfo{author}{\bibfnamefont{M.}~\bibnamefont{Fabrizio}},
  \bibinfo{journal}{Phys. Rev. B} \textbf{\bibinfo{volume}{48}},
  \bibinfo{pages}{15838} (\bibinfo{year}{1993}).

\bibitem[{\citenamefont{Khveshchenko and Rice}(1994)}]{Khveshchenko1994a}
\bibinfo{author}{\bibfnamefont{D.~V.} \bibnamefont{Khveshchenko}}
  \bibnamefont{and} \bibinfo{author}{\bibfnamefont{T.~M.} \bibnamefont{Rice}},
  \bibinfo{journal}{Phys. Rev. B} \textbf{\bibinfo{volume}{50}},
  \bibinfo{pages}{252} (\bibinfo{year}{1994}).

\bibitem[{\citenamefont{Balents and Fisher}(1996)}]{Balents1996}
\bibinfo{author}{\bibfnamefont{L.}~\bibnamefont{Balents}} \bibnamefont{and}
  \bibinfo{author}{\bibfnamefont{M.~P.~A.} \bibnamefont{Fisher}},
  \bibinfo{journal}{Phys. Rev. B} \textbf{\bibinfo{volume}{53}},
  \bibinfo{pages}{12133} (\bibinfo{year}{1996}).

\bibitem[{\citenamefont{Tsuchiizu and Furusaki}(2002)}]{Tsuchiizu2002a}
\bibinfo{author}{\bibfnamefont{M.}~\bibnamefont{Tsuchiizu}} \bibnamefont{and}
  \bibinfo{author}{\bibfnamefont{A.}~\bibnamefont{Furusaki}},
  \bibinfo{journal}{Phys. Rev. B} \textbf{\bibinfo{volume}{66}},
  \bibinfo{pages}{245106} (\bibinfo{year}{2002}).

\bibitem[{\citenamefont{Tsuchiizu and Suzumura}(2005)}]{Tsuchiizu2005}
\bibinfo{author}{\bibfnamefont{M.}~\bibnamefont{Tsuchiizu}} \bibnamefont{and}
  \bibinfo{author}{\bibfnamefont{Y.}~\bibnamefont{Suzumura}},
  \bibinfo{journal}{Phys. Rev. B} \textbf{\bibinfo{volume}{72}},
  \bibinfo{pages}{075121} (\bibinfo{year}{2005}).

\bibitem[{\citenamefont{Chudzinski et~al.}(2008)\citenamefont{Chudzinski,
  Gabay, and Giamarchi}}]{Chudzinski2008}
\bibinfo{author}{\bibfnamefont{P.}~\bibnamefont{Chudzinski}},
  \bibinfo{author}{\bibfnamefont{M.}~\bibnamefont{Gabay}}, \bibnamefont{and}
  \bibinfo{author}{\bibfnamefont{T.}~\bibnamefont{Giamarchi}},
  \bibinfo{journal}{Phys. Rev. B} \textbf{\bibinfo{volume}{78}},
  \bibinfo{pages}{075124} (\bibinfo{year}{2008}).

\bibitem[{\citenamefont{Noack et~al.}(1996)\citenamefont{Noack, White, and
  Scalapino}}]{Noack1996}
\bibinfo{author}{\bibfnamefont{R.}~\bibnamefont{Noack}},
  \bibinfo{author}{\bibfnamefont{S.}~\bibnamefont{White}}, \bibnamefont{and}
  \bibinfo{author}{\bibfnamefont{D.}~\bibnamefont{Scalapino}},
  \bibinfo{journal}{Physica C: Superconductivity}
  \textbf{\bibinfo{volume}{270}}, \bibinfo{pages}{281 } (\bibinfo{year}{1996}).

\bibitem[{\citenamefont{Khveshchenko}(1994)}]{Khveshchenko1994}
\bibinfo{author}{\bibfnamefont{D.~V.} \bibnamefont{Khveshchenko}},
  \bibinfo{journal}{Phys. Rev. B} \textbf{\bibinfo{volume}{50}},
  \bibinfo{pages}{380} (\bibinfo{year}{1994}).

\bibitem[{\citenamefont{Sedlmayr et~al.}(2011)\citenamefont{Sedlmayr, Eggert,
  and Sirker}}]{Sedlmayr2011b}
\bibinfo{author}{\bibfnamefont{N.}~\bibnamefont{Sedlmayr}},
  \bibinfo{author}{\bibfnamefont{S.}~\bibnamefont{Eggert}}, \bibnamefont{and}
  \bibinfo{author}{\bibfnamefont{J.}~\bibnamefont{Sirker}},
  \bibinfo{journal}{Phys. Rev. B} \textbf{\bibinfo{volume}{84}},
  \bibinfo{pages}{024424} (\bibinfo{year}{2011}).

\bibitem[{\citenamefont{Sedlmayr et~al.}(2012)\citenamefont{Sedlmayr, Ohst,
  Affleck, Sirker, and Eggert}}]{Sedlmayr2012a}
\bibinfo{author}{\bibfnamefont{N.}~\bibnamefont{Sedlmayr}},
  \bibinfo{author}{\bibfnamefont{J.}~\bibnamefont{Ohst}},
  \bibinfo{author}{\bibfnamefont{I.}~\bibnamefont{Affleck}},
  \bibinfo{author}{\bibfnamefont{J.}~\bibnamefont{Sirker}}, \bibnamefont{and}
  \bibinfo{author}{\bibfnamefont{S.}~\bibnamefont{Eggert}},
  \bibinfo{journal}{Phys. Rev. B} \textbf{\bibinfo{volume}{86}},
  \bibinfo{pages}{121302} (\bibinfo{year}{2012}).

\bibitem[{\citenamefont{Sedlmayr et~al.}(2013)\citenamefont{Sedlmayr, Adam, and
  Sirker}}]{Sedlmayr2013}
\bibinfo{author}{\bibfnamefont{N.}~\bibnamefont{Sedlmayr}},
  \bibinfo{author}{\bibfnamefont{P.}~\bibnamefont{Adam}}, \bibnamefont{and}
  \bibinfo{author}{\bibfnamefont{J.}~\bibnamefont{Sirker}},
  \bibinfo{journal}{Phys. Rev. B} \textbf{\bibinfo{volume}{87}},
  \bibinfo{pages}{035439} (\bibinfo{year}{2013}).

\bibitem[{\citenamefont{Giamarchi and Schulz}(1988)}]{Giamarchi1988}
\bibinfo{author}{\bibfnamefont{T.}~\bibnamefont{Giamarchi}} \bibnamefont{and}
  \bibinfo{author}{\bibfnamefont{H.}~\bibnamefont{Schulz}},
  \bibinfo{journal}{J. Phys. France} \textbf{\bibinfo{volume}{49}},
  \bibinfo{pages}{819} (\bibinfo{year}{1988}).

\bibitem[{\citenamefont{Chang and Affleck}(2007)}]{Chang2007}
\bibinfo{author}{\bibfnamefont{M.-S.} \bibnamefont{Chang}} \bibnamefont{and}
  \bibinfo{author}{\bibfnamefont{I.}~\bibnamefont{Affleck}},
  \bibinfo{journal}{Phys. Rev. B} \textbf{\bibinfo{volume}{76}},
  \bibinfo{pages}{054521} (\bibinfo{year}{2007}).

\bibitem[{\citenamefont{Garate and Affleck}(2010)}]{GarateAffleck}
\bibinfo{author}{\bibfnamefont{I.} \bibnamefont{Garate}} \bibnamefont{and}
  \bibinfo{author}{\bibfnamefont{I.}~\bibnamefont{Affleck}},
  \bibinfo{journal}{Phys. Rev. B} \textbf{\bibinfo{volume}{81}},
  \bibinfo{pages}{144419} (\bibinfo{year}{2010}).

\bibitem[{\citenamefont{Juttner and Klumper}(2010)}]{JuttnerKlumper}
\bibinfo{author}{\bibfnamefont{G.} \bibnamefont{J\"uttner}},
  \bibinfo{author}{\bibfnamefont{A.}~\bibnamefont{Kl\"umper}}, \bibnamefont{and}
\bibinfo{author}{\bibfnamefont{J.}~\bibnamefont{Suzuki}},
  \bibinfo{journal}{Nucl. Phys. B} \textbf{\bibinfo{volume}{522}},
  \bibinfo{pages}{471} (\bibinfo{year}{1998}).

\bibitem[{\citenamefont{Meden and Schoenhammer}(1992)}]{Meden1992}
\bibinfo{author}{\bibfnamefont{V.}~\bibnamefont{Meden}} \bibnamefont{and}
  \bibinfo{author}{\bibfnamefont{K.}~\bibnamefont{Schoenhammer}},
  \bibinfo{journal}{Phys. Rev. B} \textbf{\bibinfo{volume}{46}},
  \bibinfo{pages}{15753} (\bibinfo{year}{1992}).

\bibitem[{\citenamefont{Voit}(1995)}]{Voit1995}
\bibinfo{author}{\bibfnamefont{J.}~\bibnamefont{Voit}}, \bibinfo{journal}{Rep.
  Prog. Phys.} \textbf{\bibinfo{volume}{58}}, \bibinfo{pages}{977}
  (\bibinfo{year}{1995}).

\bibitem[{\citenamefont{Braunecker et~al.}(2012)\citenamefont{Braunecker, Bena,
  and Simon}}]{Braunecker2012}
\bibinfo{author}{\bibfnamefont{B.}~\bibnamefont{Braunecker}},
  \bibinfo{author}{\bibfnamefont{C.}~\bibnamefont{Bena}}, \bibnamefont{and}
  \bibinfo{author}{\bibfnamefont{P.}~\bibnamefont{Simon}},
  \bibinfo{journal}{Phys. Rev. B} \textbf{\bibinfo{volume}{85}},
  \bibinfo{pages}{035136} (\bibinfo{year}{2012}).

\bibitem[{\citenamefont{Schuricht et~al.}(2013)\citenamefont{Schuricht,
  Andergassen, and Meden}}]{Schuricht2013}
\bibinfo{author}{\bibfnamefont{D.}~\bibnamefont{Schuricht}},
  \bibinfo{author}{\bibfnamefont{S.}~\bibnamefont{Andergassen}},
  \bibnamefont{and} \bibinfo{author}{\bibfnamefont{V.}~\bibnamefont{Meden}},
  \bibinfo{journal}{Journal of Physics: Condensed Matter}
  \textbf{\bibinfo{volume}{25}}, \bibinfo{pages}{014003}
  (\bibinfo{year}{2013}).

\end{thebibliography}
\end{document}